\documentclass[fleqn,usenatbib]{mnras}

\usepackage[T1]{fontenc}
\usepackage{ae,aecompl}

\usepackage{graphicx}   
\graphicspath{{AASTEX//}}

\usepackage{amsmath}    
\usepackage{amssymb}    

\usepackage{xfrac}
\usepackage{float}
\usepackage{physics}
\usepackage{amssymb}
\usepackage{savesym}
\savesymbol{tablenum}
\usepackage{rotating}
\usepackage[section]{placeins}
\usepackage[noabbrev]{cleveref}
\usepackage{subfigure}
\usepackage{gensymb}
\usepackage{siunitx}

\newcommand{\HI}{{H}{I}}

\newcommand{\kms}{km s$^{-1}$}
\newcommand{\dunit}{m \textsuperscript{-3}}
\restoresymbol{SIX}{tablenum}

\newcommand{\tus}[1]{$_{\text{2}}$}
\newcommand{\hh}{$H_2$}

\usepackage{natbib}
\usepackage{url}
\usepackage{appendix}
\usepackage{hyperref}

\title[Inclined Gas Disc Collisions]{Splash Bridge Models of Inclined, Gas-Rich, Direct Galaxy Collisions}

\author[Yeager and Struck]{
Travis R. Yeager,$^{1}$\thanks{E-mail: yeagerastro@gmail.com (TRY)}
and Curtis Struck,$^{1}$
\\
$^{1}$Iowa State University, Department of Physics and Astronomy, Ames, IA, US\\
}

\date{Accepted Jan 10, 2020. Received Nov 04, 2019; in original form Dec 18, 2020.}

\pubyear{2019}

\begin{document}
\label{firstpage}
\pagerange{\pageref{firstpage}--\pageref{lastpage}}
\maketitle

\begin{abstract}
Splash bridges are formed from the direct inelastic collision of gas-rich galaxies. Recent multi-wavelength observations of the Taffy galaxies, UGC 12914/15, have revealed complicated gas structures in the bridge.  We have upgraded the sticky particle simulation code of \citet{yeager19} by adding: the ability to adjust the relative inclination of the gas discs, the ability to track cloud-cloud collisions over time, and additional cooling processes.  Inclination effects lead to various morphological features, including filamentary streams of gas stripped from the smaller galactic disc.  The offset of disc centres at impact determines whether or not these streams flow in a single direction or multiple directions, even transverse to the motion of the two galaxies.  We also find that, across many types of direct collision, independent of the inclination or offset, the distributions of weighted Mach numbers and shock velocities in colliding clouds relax to a very similar form.  There is good evidence of prolonged turbulence in the gas of each splash bridge for all inclinations and offsets tested, as a result of continuing cloud collisions, which in turn are the result of shearing and differentially accelerated trajectories.  The number distribution of high velocity shocks in cloud collisions, produced in our low inclination models, are in agreement with those observed in the Taffy Galaxies with ALMA, \citep{almataffy2019}.

\end{abstract}



\section{Introduction: Splash Bridges}

\indent Galaxy interactions play an important role in galaxy evolution, morphology and star formation.  The effects of gravitational interactions between galaxies that produce tidal tails and ring waves are distinctly different than the direct collisions between gas elements that produce splash bridges after a direct disc-disc collision.  

\indent For example, the Taffy galaxy system, UGC 12914/15, contains a multi-phase gas bridge of material stretched between the two stellar discs.  There are over 25 years of detailed observations of the Taffy ring galaxies, which have apparently recently undergone a direct collision, \citet{condon93,braine03,gao03,peterson12,appleton15,almataffy2019}.  It is likely that splash bridges will be found in many galaxy systems similar to the Taffy, especially the colliding ring galaxies.  Arp 143, \citet{Arp143H22009,Arp1432014} and Arp 194, \citet{Arp1942003}, are further examples of possible splash bridge systems.  HI observations are not available for many of these systems and in some systems one of the galaxies may be gas poor (e.g. Arp 147 \citet{fog11} and the Lindsay-Shapley ring, \citet{lindshapring2012}).  Moreover, in the impacted regions the gas column densities of the two discs may be very different, e.g., the vertical column density of one disc versus whole in-plane column density of the other in collisions with large relative inclinations. Thus, the gas from the sliced disc essentially accretes onto the other galaxy, with little splashed into the bridge (see \citet{struck97simulations}). Arp 271, UGC 7125/26, and ESO 138 - IG029 (the 'Sacred Mushroom') may be examples of slicing impacts based on the fragmentary HI bridges in the HI Rogues Gallery \citep{hibbard01}. 

\indent The HI Rogues Gallery does contain a few ring systems with HI bridges, including: Arp 284, Arp 298, the Cartwheel, and VII Zw 466. Most of these probably involved somewhat tilted (partially slicing) impacts, whereas the Taffy impact was evidently nearly face-on for both galaxies. Arp 284, in particular, probably had a somewhat off-centre, partial fly-by encounter \citep{smithstruck97}. Given the extended HI discs of many galaxies, still more off-centre impacts may generate splash bridges, but not a ring galaxy.  

\indent \citet{davies08} outline the detection of a dark galaxy candidate, VIRGOHI21, which could be a relic of a splash bridge.  Interestingly, the \HI{} observations of VIRGOHI21 reveal an \HI{} galactic disc with a mass of \SI{2e7}{M\textsubscript{\(\odot\)}} embedded in the \SI{2e8}{M\textsubscript{\(\odot\)}} tidal bridge.  Stars are not observed in this object; a lower limit for mass to luminosity ratio was found to be \SI{e6} {M\textsubscript{\(\odot\)}}/{L\textsubscript{\(\odot\)}}, which the authors note is significantly higher than a typical galactic value of 50 or less.  Further discussion of splash bridges can be found in  \citet[Paper I]{yeager19}.

\indent The observational resolution required to begin to see detailed structure in Taffy-like bridges has been achieved with ALMA.  Results of \HI{} observations of the Taffy reveal clumps and filaments across the splash bridge.  This is in close accord with the results presented in this paper.  

\indent The interstellar medium (ISM) of our initial discs is modelled as five discrete gas phases, with initial temperatures and densities typical of the phases in a gas-rich Sc galaxy.  Each initial disc is divided into a uniform grid of cells, and each grid cell is assumed to contain a spherical gas element.  Each of these gas elements is treated as a test particle (i.e., no self-gravity) and will be referred to in this paper as a 'cloud'.  The initial temperature and density of each cloud is assigned based on its position in the disc and its phase.  The initial distribution of each phase of ISM in the initial discs can have an strong influence on the resulting bridge evolution and structure, since the cloud density will change its inertial resistance to the gas ram pressure in a collision.  To better understand the evolution of the ISM involved in galaxy collisions over time scales of \SI{e8} yr, we have simulated collisions between gas-rich discs across several inclinations at both small (500 pc) and large (10 kpc) impact offsets of the galaxy centres.  We have found that some splash bridge characteristics are indicative of specific collision parameters.  In the disc-disc collisions, when the centre of a cloud moves within the radius of another cloud a collision is recorded.  Every collision between gas clouds is assumed to create a shock wave in both gas clouds, heating and compacting them.  The shock crossing time is estimated from the cloud's size and shock velocity, after which cooling is allowed to begin within the cloud.  The cooling that is applied to each cloud depends on its current temperature and density.  These simple prescriptions allow us to track cloud temperature and density populations over time as the galaxies interact.  We are also able to study the distribution of shock velocities between in clouds of each phase of ISM over time.  This reveals the turbulence, driven by the kinematics as described below, occurring in the gas of these splash bridges.

\indent The role turbulence may play in triggering or quenching star formation in these bridges remains an open question.  Across all types of collision we find continued cloud collisions in the model splash bridges up to the end of the simulation, which is taken to be about 120 Myr after the galaxies nearest approach.  This prolonged turbulence is enough to slow the cooling of the hottest phases of gas, as well as heat a small fraction of colder gas to line emission temperatures or greater.  In some cases the rate of collisions begins to grow 50 Myr after the nearest approach, causing a growth in the number of gas clouds hot enough for X-ray emission.

\section{The Model}
The terms G1 and G2 will be used in this paper to refer to the model galaxies.  G1 is the more massive (like UGC 12914 in the Taffy system) galaxy and G2 (like UGC 12915) the less massive.

\subsection{Numerical Methods}

\indent To investigate the effect disc inclination has on splash bridges we extended the inelastic particle code employed in, \citet{yeager19}.  A complete description of the basic method and initial conditions is given in section 2 of \citet{yeager19}. The following discussion focuses on changes to that code made for the present simulations. 

\subsubsection{Conditions for Cloud-Cloud Collisions}

\begin{figure}
   \label{fig:cloudcollision}
  \includegraphics[width=0.4\textwidth]{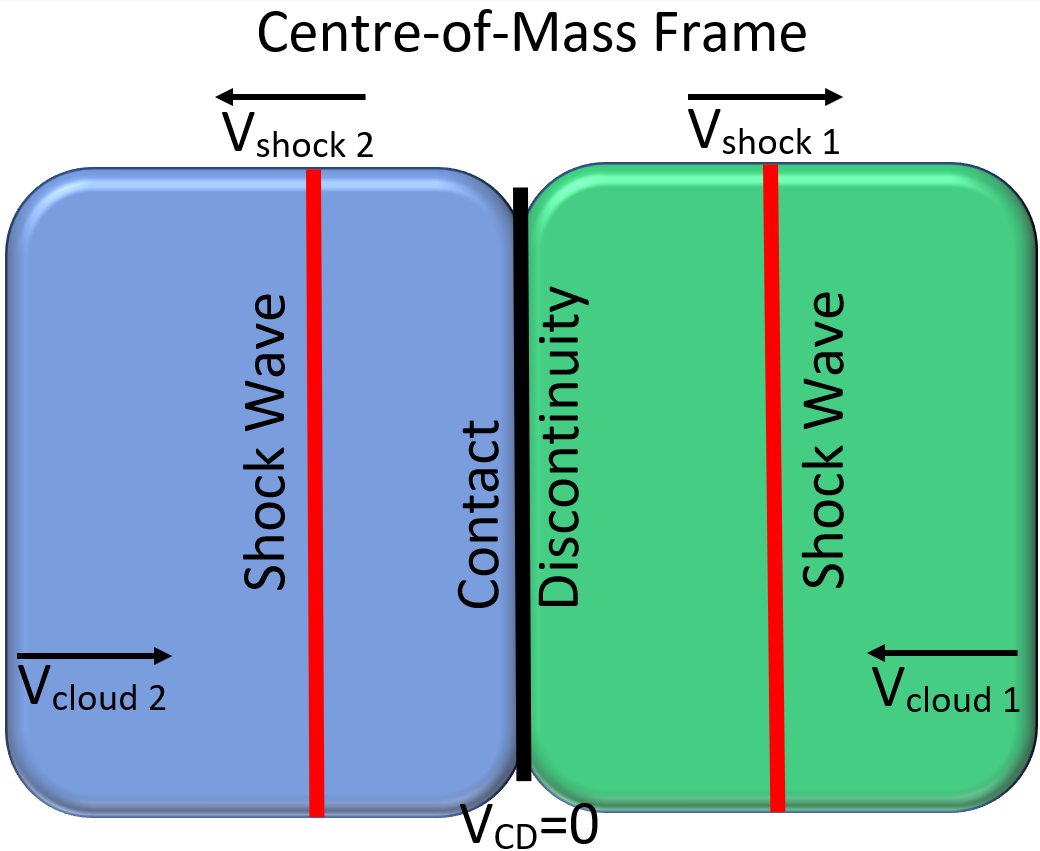}
 \caption{Diagram of a single collision between two gas elements.  Gas left behind the shock is at rest with respect to the two cloud's centre-of-mass. $V_{shock}$ is the velocity of the shock in the COM frame.  $V_{cloud}$ is the velocity of each cloud in the COM frame.}%
\end{figure}

\indent Cloud collisions are detected with a nearest neighbour search after every time step.  This detection frequency is a departure from the previous version of this code, but is necessary since introducing large disc-disc inclinations prevents us from using a single grid to define colliding gas clouds.  Collisions between gas clouds are handled as inelastic momentum conserving collisions as in the previous face-on models.  \Cref{fig:cloudcollision} is a diagram of an individual collision.  The gas clouds are depicted as cube-like in the shock diagram for clarity in showing the contact discontinuity and shocks.  The shocks are produced at the contact discontinuity, a boundary where the gas clouds make first contact.  From this discontinuity two shock waves are produced, each propagating back through its parent cloud.  

\indent It is possible for multiple gas clouds to find themselves within each others radii over a given time step. However, for computational efficiency, we count only the nearest gas clouds as colliding.  To make clear exactly what is occurring consider the following example.  Label four gas clouds: A, B, C, and D.  At a single time step cloud A finds that all three gas clouds B, C and D are within its collision radius.  The code determines that cloud C is closest to the centre of A, and a collision is recorded between A and C.  When it comes to the nearest neighbour search on cloud B, it may still result that A is its closest neighbour, and then a collision between B and A is recorded in the same time step.  Since it was found that A was closest to C no new collision would be recorded, but if D was found closest to C, there would be a second collision for C in the same time step.  When it computing cooling later, only the strongest shock is considered in each time step (e.g., for cloud A, which is involved in multiple shocks, in this example).  

\indent Every cloud is given an initial 'collision radius' that does not change with time, equal to half the set resolution of the simulation, which for all cases here is 100 pc, unless specified otherwise.  A second 'cooling radius' is updated for each cloud as the cloud cools or adiabatically expands.  This allows us to computationally separate the cooling processes from the kinematics.  It also keeps the amount of mass overlapping in each collision  similar over time.  If the centre of one cloud is within another gas cloud radius a collision may occur.  A second requirement for collision is that only gas clouds with relative velocities above 40 \kms{} are considered.  The relative velocity limit prevents gas clouds from continuously registering a collision in each time step after they have collided.  It is also important to prevent each disc from triggering gas cloud collisions due to rotation prior to the disc-disc collision as this would warm our coldest phases of ISM.  By assuming each disc is an Sc-type-disc it is also assumed there are the required processes to allow a cold phase of ISM exist.  Clouds colliding with relative velocities below 40 \kms{} will experience a maximum heating of a few thousand degrees K, which will have little affect much of the medium in this run.

\indent We estimate that the strongest pressure gradient forces, which are not computed correctly by the sticky particle algorithm could alter trajectories of a small number of particles by less than 1.0 kpc over the duration of a 150 Myr run.

\indent To test the effect particle resolution has on our model, we have completed runs with particle resolutions of 70, 140, 200, and 500 pc.  The results, shown in \cref{fig:resolutions1,fig:resolutions2}, converge and are consistent across the resolutions tested.  

\indent These simple collision rules allow for rapid computations, as do the cooling functions considered in the next subsection. They cannot accurately model hydrodynamic forces in the long term, but do provide an efficient, quantitative modeling of the thermo-hydrodynamic processes over a modest timescale, e.g., over the time of bridge formation and the first couple of collisions for a typical cloud.

\subsubsection{Cloud Cooling}

\indent Shocks are calculated for every collision, but are only applied if they produce a post shock temperature greater than their cloud's temperature at half the crossing time.  If a shock is strong enough the cloud temperature is increased to the post-shock temperature and the density and radius are updated corresponding to the shock jump conditions when the half crossing time is completed.  After a shock has completed half a crossing time, cooling and heating processes begin within the cloud.

\indent The cooling calculation has been changed from an evaluation across the dynamical time step to one using the Matlab ode23 solver to compute the equation for the temperature rate of change. The latter uses a variable time step integration over the duration of each dynamical time step.  By moving from a fixed cooling time step a more careful analysis of $T < 20,000$K cooling can be done.  Previously we only included CII fine structure cooling and a uniform UV heating at $T < 20,000$K.  This leads to gas that cools to densities above the CII critical density for collisional deexcitation, and then hanging up at $T = 20,000$K.  

\indent In the present code a form of dust grain cooling has been included.  The radius of the dust grains ($a$) is allowed to evolve via sputtering as per \cref{eq:sputteringradius} from \citet{drainephysics},

\begin{equation}
    \label{eq:sputteringradius}
    \dv{a}{t} = - \frac{10^{-6}}{(1 + T_{6}^{-3})} \frac{n_{h}}{(cm^{-3})}\ \frac{\si{\micro \meter}}{yr}.
\end{equation}


If grains sputter below radius of 1.0 nm, dust grain cooling is halted.  When a dust grain collides with gas atoms most of the kinetic energy is transferred to the dust grain.  Dust grains cool extremely fast compared to galactic time scales, so it is assumed that the gas instantly radiates $0.8kT$ Joules of heat after each dust grain collision.  The dust collision rate ($\dv{N_{col}}{t}$) is given by,  

\begin{equation}
    \label{eq:colrate}
    \dv{N_{col}}{t} = n_{h} n_{dust} \pi a^2 v_{prob},
\end{equation}

and

\begin{equation}
    \label{eq:dustrate}
    \dv{T}{t} = -.8 \dv{N_{col}}{t} T.
\end{equation}

\noindent Here $v_{prob}$ is the most probable velocity of a single ISM particle, $n_h$ is the number density of the ISM, $n_{dust}$ is the number density of dust grains, $a$ is the current grain radius and $T$ is the temperature of the local ISM.  

\indent As in Paper I cooling by adiabatic expansion, hydrogen and helium line cooling, Bremsstrahlung and fine structure is also applied. To calculate the cooling function for optically thin cooling by hydrogen and helium emission, as well as Bremsstrahlung, we use the following equation from  \citet{wang15}, 

\begin{equation} \Lambda (T) = \frac{{x_1} T^{x_2} + ({x_3} T)^{x_4} ({x_5} T^{x_6} + x_7 T^{x_7})}{1 + ({x_3} T)^{x_4}} + {x_9} T^{x_{10}}.
\label{eq:coolingfunction}
\end{equation}

\noindent This equation is fit with the constants from \cref{tab:cfconsts}). This function is valid at temperatures of \SI{2e4} to \SI{e10} K and densities of less than \SI{e18} m\textsuperscript{-3}, which is well beyond the range required for our models.

\begin{table}
\caption{Cooling function constants for \cref{eq:coolingfunction}.}
\label{tab:cfconsts}
\begin{tabular}{| c | c | c | c |}
\hline
Constant & Value & Constant & Value \\
\hline
$x_1$ & \SI{4.86567e-13} & $x_2$ & -2.21974 \\
$x_3$ & \SI{1.35332e-5} & $x_4$ & 9.64775 \\
$x_5$ & \SI{1.11401e-9} & $x_6$ & -2.66528 \\
$x_7$ & \SI{6.91908e-21} & $x_8$ & -0.571255 \\
$x_9$ & \SI{2.45596e-27} & $x_{10}$ & 0.49521 \\
\hline
\end{tabular}
\end{table}

\indent Following the shock compression, cooling by adiabatic expansion occurs as the cloud expands spherically at the speed of sound (c) given by \cref{eq:soundc}.  A $\gamma$ of $\frac{5}{3}$ is used and the molecular weight $\mu$ is calculated for a mixture of hydrogen and He gas at a 3 to 1 ratio, $m_h$ is the mass of hydrogen. Then, 

\begin{equation} c = \sqrt{\gamma \frac{k_{B} T}{\mu m_{h}}}
\label{eq:soundc},
\end{equation}

\begin{equation} \dv{T_{adb}}{t} = \frac{-3(\gamma-1) c T}{R}.
\label{eq:workbyadb}
\end{equation}

An intergalactic pressure of \SI{e2} \dunit K is assumed for every cloud, \citet{Nicastro2018}.  If a cloud is below this pressure threshold adiabatic expansion will halt.  

\indent CII fine structure cooling is approximated by, 

\begin{equation} \dv{T_{CII}}{t} = \frac{-2(\SI{3.154e-32}) n^{2} }{3*k_{b}*n_{i}}
\label{eq:fine}
\end{equation}

\noindent this is estimated from figure 30.1 and the critical density from table 17.1 of \citet{drainephysics}.  This cooling is set to zero above number densities (n) of \SI{2e9} \dunit.

\indent A constant rate of UV heating is considered.  This rate is approximately $\frac{1}{3}$ of the CII cooling rate, $\Lambda_{UV} = \SI{9.461e-33} J m^{3} s^{-1}$ also taken from \citet{drainephysics}.

\indent  We have looked across the parameter space of temperatures 150-\SI{e7} K and densities \SI{e2}-\SI{e12} \dunit{} to find where gas cloud cooling may be balanced by the constant UV heating.  We find that gas clouds are unable to move through the parameter region of densities greater than \SI{2e9} \dunit{} (CII cooling limit) and temperatures less than 20,000 K (line cooling limit). The adiabatic cooling rate for a 100 pc cloud is about two orders of magnitude less than the UV heating rate.  Since the adiabatic cooling rate scales linearly with cloud radius, the two rates are approximately equal around cloud radii of 1 pc.  Initially less than $10\%$ of the ISM in each galaxy is within the region where UV heating is greater than the cooling, though as shocks from collisions compress and heat gas clouds, densities of \SI{e7} \dunit are pushed into this non-cooling region.

\subsection{Initial Conditions}

\begin{figure}
  
  \includegraphics[width=0.4\textwidth]{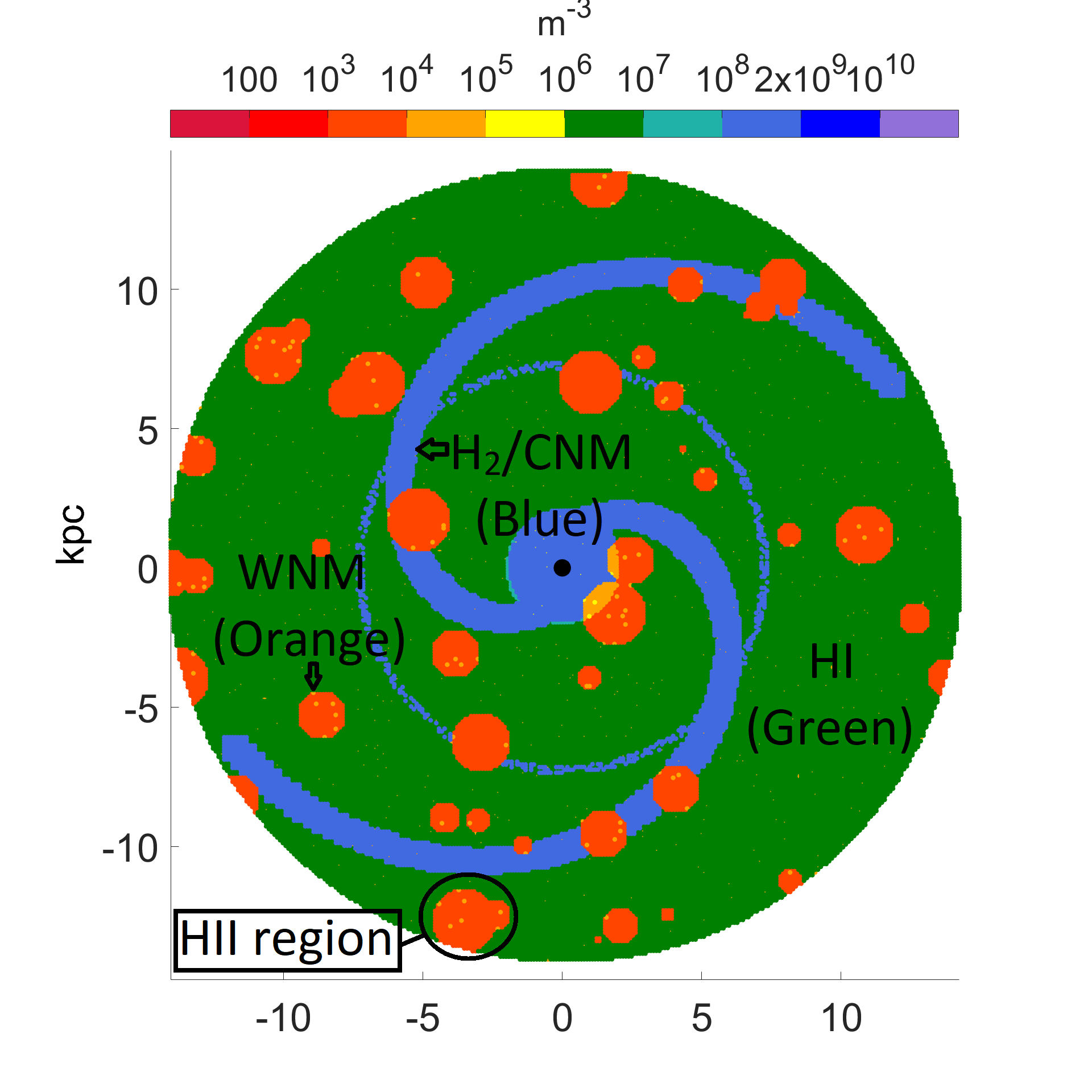}
 \caption{Example initial ISM distributions and densities.  The circular regions in red are the HII hot gas 200 pc to 2 kpc in size, in blue the spiral arms contain \hh{} material and green makes up the most abundant component of neutral Hydrogen.  Not easily seen are 100 pc gas clouds of light blue and orange scattered randomly through the disc, representing the cold neutral and warm neutral ISM phases. }%
 \label{fig:initialdisc}
\end{figure}

\indent The discs are initialized with the same parameters as in the face-on models of Paper I except as noted in this section.  An example of an initial disc is shown in \cref{fig:initialdisc}.  The initial separation between galaxy centres (z) is set to -15 kpc for every run. In all cases, the smaller G2 galaxy is placed below G1 in the vertical (z) direction.  An initial velocity for G2 is set to 400 $km/s$ in the z direction toward G1.  The disc plane of galaxy G2 is inclined relative to the x-y plane by different amounts in different runs.  The disc gravitational potential of G2 is also inclined to match the initial gas disc of G2.  

\indent Each run starts approximately 25 Myr before the galactic centres reach closest approach.  This allows for gravitational effects to begin warping the discs prior to the collision.  Initially, the two discs are given: a separation in x,y,z, a velocity in the z direction, and an inclination measured from the x-y plane about the x-axis (same axis of offset), unless otherwise specified.  The offset is defined as how far apart the galactic centres are initially placed.  

\indent The initial phases of \hh, cold neutral \HI, warm \HI, hot hydrogen and fully ionized hydrogen are given initial temperatures of 50, 100, 5000, \SI{5e5}, \SI{5e6} Kelvin, respectively.  These values and corresponding densities are selected to maintain pressure equilibrium through the initial disc.  The added dust grains are treated as spheres with an initial radius $a_i = 100 nm$.  The number density of dust ($n_d$) is set so that the local dust mass is equal to 1\% of the mass of the local ISM.  Dust number density is not changed within each cloud element; cooling effectiveness is reduced as the grain size decreases through sputtering.  These phases of ISM do not evolve through heating or cooling prior to their first collision.  The ISM is initialized to represent a typical Sc galaxy, so it is assumed there are internal processes in each phase that maintain the relative temperatures and densities over the approximate 10 Myr before they experience a collision.

\indent The distribution of hot diffuse gas has been changed from the models of Paper I. Previously, the hot gas was dispersed randomly throughout the disc as individual 100 pc gas `clouds'.  While these gas clouds could by chance be near each other, in actuality there were few regions of adjacent hot gas.  Now about 50 regions have been chosen to be filled in with hot diffuse gas, each randomly placed with a randomly chosen diameter of 200 pc to 2 kpc.  This was done to better represent star forming regions and the holes they produce in gas-rich galaxies.  \Cref{fig:initialdisc} shows a typical initial G1 disc; the hot diffuse gas regions are shown in red.  

\flushbottom
\section{Results}

\indent We have computed several dozen simulations of various resolutions, relative rotations and inclinations.  After some analysis we chose eight runs, four with 500 pc initial x-offsets and four with 10 kpc offsets, to be representative of the inclination and offset parameter space.  In \cref{sec:bridges} we will discuss what is unique about each of the eight runs, then in \cref{sec:distributions} we take a closer look at the shock and Mach number distributions in colliding clouds, both in total over the runs, and as a function of time for some of the collisions.


\subsection{Discussion of Morphology and Bridge Characteristics}
\label{sec:bridges}
\indent In each of the model figures below the system's age is given in the top right, and the line-of-sight direction in the bottom right.  The age is measured from the time of nearest approach of the galaxy centers.  The line-of-sight is given by an altitude and azimuth which are the Matlab figure view coordinates.  The azimuth is the angle of rotation around the z-axis, as measured from the negative y-axis. Increasing this angle corresponds to counterclockwise rotation about the z-axis when viewing the x-y plane from above.  The elevation is the minimum angle between the line-of-sight and the x-y plane. Increasing the elevation from -90 to 90 degrees corresponds to a rotation from the negative z-axis to the positive z-axis.

\indent Since we have integrated these systems over a time of 150 Myr it is not possible to show figures of the entire evolution.  Videos of these systems evolving will be added to a YouTube account which may be found at: \url{bit.ly/splashbridges}.  In the figures below we show every system at an age of 30 Myr, which is the estimated age of the splash bridge in the Taffy system.  For most of the runs we also include figures at about 90 Myr after the collision.  

\indent In all of the models we find that the temperature of the bridge reaches post-impact values of order \SI{e7} K, due to the main disc collision.  Inclination determines whether the initial heating occurs across each galaxy rapidly or is spread over the time interval it takes different parts of G2 to cross G1.  For low inclinations the gas experiences less turbulence and takes between 20 - 30 Myr to cool.  Some parts of each bridge will remain hot due to either continued collisions or the fact that the gas is diffuse enough that the cooling time is greater than about 100 Myr.  Reheating of the ISM can been seen in every model as gas falls back onto the central region of the larger G1.  The presence of substantially bigger 'holes' in the initial discs of the present models allows the creation of similar size clumps of gas that are only displaced a kiloparsec or so from their host galaxy.

\indent Another effect occurs in all of our splash bridges with inclinations $45\degree{}$ and greater.  Visible in the figures, a gradient of cloud temperature and densities can be seen through the bridge.  On one side are gas clouds of density \SI{e7} \dunit{} that remain over \SI{e5} K and on the other, cooled gas clouds with densities greater than \SI{e10} \dunit.  This boundary is more obvious in some runs than others and does not always occur along the same line of sight.  The rotation of G2 seems to be the cause of this divide, between gas clouds.  Rotation velocities in each galaxy peak between 200-300 \kms{} whereas the centers of each galaxy are traveling around 550 \kms{} as they pass one another.  Gas clouds in G2 rotating with a positive z-direction component, are propelled into G1 as the discs collide, and experience a pile up between gas clouds as they are accelerated back toward G1 after the collision.  The decreased distance between gas clouds causes many more collisions to occur after the main disc collision, preventing these gas clouds from cooling.  Whereas gas clouds with a negative z rotation velocity component are left traveling away from G1 with slower velocities.  The distance between gas clouds grows after the discs collide.  The rarefaction of the inter-gas cloud distances prevents continued turbulence which allows the gas clouds to cool.  

\indent Initially very hot and dense gas clouds will destroy their dust grains before line cooling can bring the cloud below 20,000 K.  By the time the cloud reaches 20,000 K it will have compressed to a density well beyond the critical density for CII fine structure cooling to kick in.  This causes the cloud to thermally "hang" since there are no longer pathways to cool the gas cloud.  Recall that the critical maximum density for CII collisional deexcitation is set at \SI{2e9} \dunit.  Thus, a cloud heated to a temperature of \SI{1e6} K will need to have a density of less than \SI{4e7} \dunit{} to allow CII cooling below 20,000 K, assuming little to no dust is left by this time.

\indent As the discs become more inclined to one another the rotation of each plays a weakened role in determining the net collision speeds.  Lower collision speeds have an obvious affect on reducing the maximum shock velocities seen between gas clouds.  Inclination also correlates with an increase in the prolonged turbulence in the gas after the collision.  The effect of inclination on a counter rotating case is less significant than in equal sense rotation collision. That is, in equal sense cases the similar rotation in the two discs provides little increase in relative collision speeds beyond the relative impact speed.

\subsubsection{Collisions with a \texorpdfstring{$20\degree$}{} Inclination}

\begin{figure*}
     \begin{center}
                \subfigure[]{%
          \includegraphics[width=0.4\textwidth]{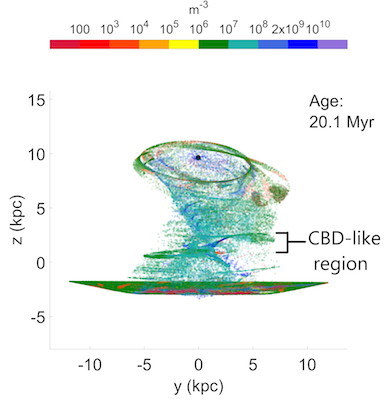}
        }
                \subfigure[]{%
          \includegraphics[width=0.4\textwidth]{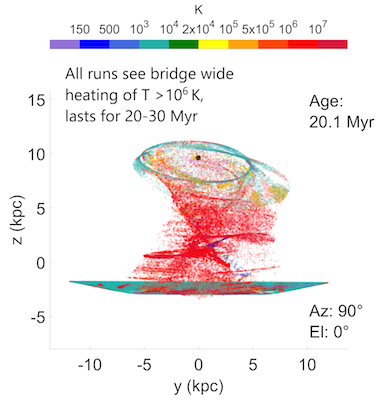}
        }\\ 
                \subfigure[]{%
          \includegraphics[width=0.4\textwidth]{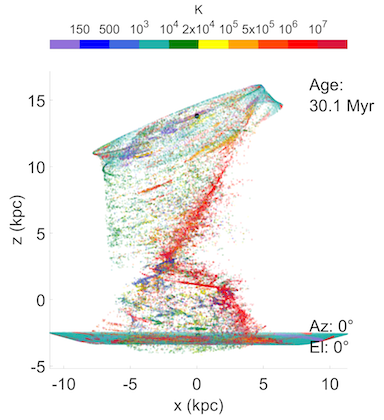}
        }
                \subfigure[]{%
          \includegraphics[width=0.4\textwidth]{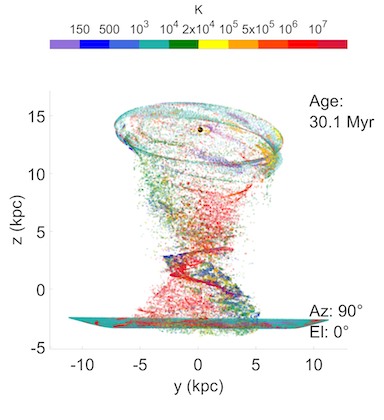}
        }\\ 
\caption{Edge-on views of the density (panel a) and temperatures (panels b, c, and d) in a model with a $20\degree$ inclination and a 500 pc initial offset.} %
\label{fig:500pc20tilt1}
    \end{center}
\end{figure*}

\begin{figure*}
     \begin{center}
                \subfigure[]{%
          \includegraphics[width=0.4\textwidth]{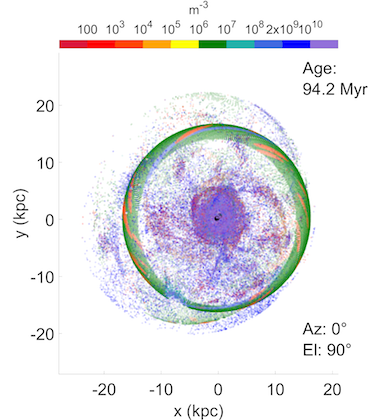}
        }
                \subfigure[]{%
          \includegraphics[width=0.4\textwidth]{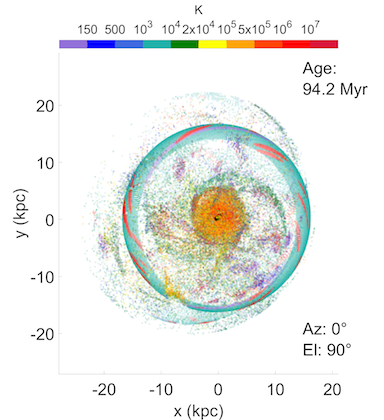}
        }\\ 
                \subfigure[]{%
          \includegraphics[width=0.4\textwidth]{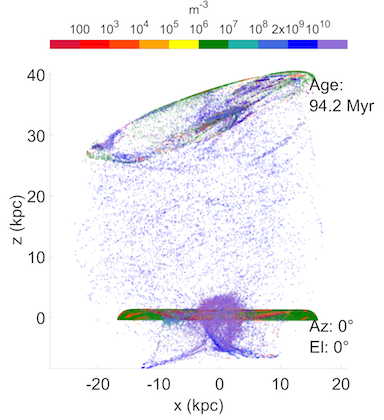}
        }
                \subfigure[]{%
          \includegraphics[width=0.4\textwidth]{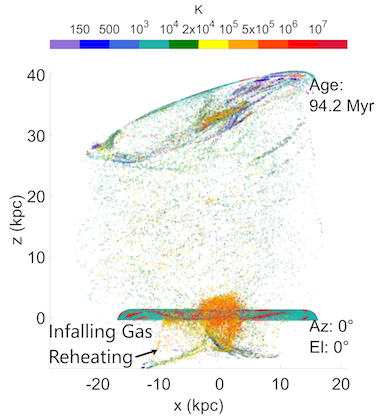}
        }\\ 
    \end{center}
\caption{Face-on views (top row) and edge-on views of the model with a $20\degree$ inclination and a 500 pc offset, at a late time. Density is shown in panels a) and c), and temperature in panels b) and d).}%
\label{fig:500pc20tilt2}
\end{figure*}

\Cref{fig:500pc20tilt1} shows a collision of two discs inclined at 20$\degree$.  In frame a) a central bridge disc (CBD)-like structure similar to the face-on models in Paper I forms.  However this is less flat and is broken apart early on, after about 25 Myr.  Rather than a single CBD with two spiral arms wound above and below it as in Paper I, as a result of multiple collisions, more gas is spread through the bridge in a three dimensional hourglass-like shape.  By 30 Myr the CBD region of gas has lost most of its disk-like form, and has begun falling back onto G1.  \Cref{fig:500pc20tilt2} shows the system about 94 Myr after the discs have collided.  In the face on views, frames a) and b), we can see the gas of G1 expanding outward in a ring.  Frames c) and d), show the system edge-on and highlight a hot bulge of gas created around the center of G1 from the infall of bridge material.  As the gas falls back onto G1, it remains heated at temperatures of order \SI{e5} K due to frequent collisions.

\begin{figure*}
     \begin{center}
                \subfigure[]{%
          \includegraphics[width=0.4\textwidth]{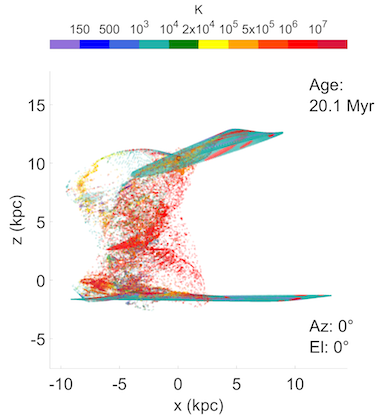}
        }
                \subfigure[]{%
          \includegraphics[width=0.4\textwidth]{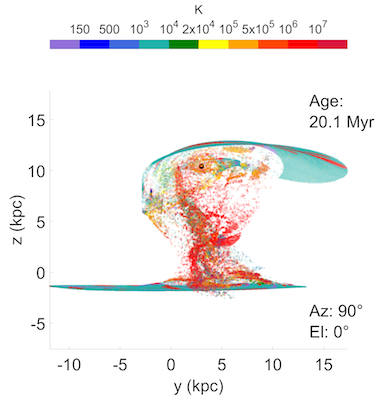}
        }\\ 
                \subfigure[]{%
          \includegraphics[width=0.4\textwidth]{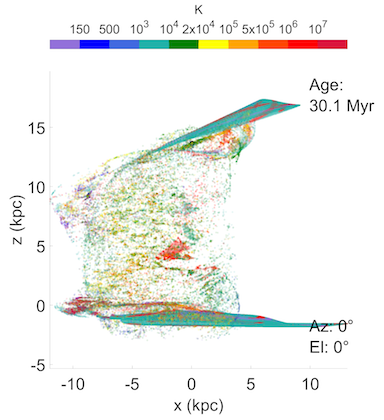}
        }
                \subfigure[]{%
          \includegraphics[width=0.4\textwidth]{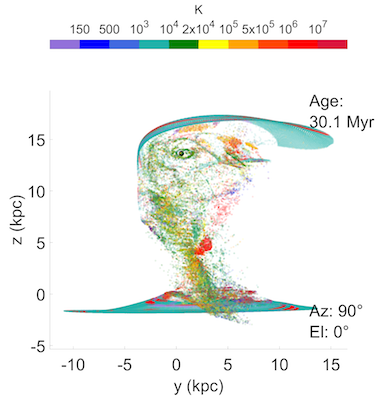}
        }\\ 
    \end{center}
\caption{Edge-on views of the temperature at two times in a model with a $20\degree$ inclination and a 10 kpc offset.}%
\label{fig:10kpc20tilt1}
\end{figure*}

\begin{figure*}
     \begin{center}
                \subfigure[]{%
          \includegraphics[width=0.4\textwidth]{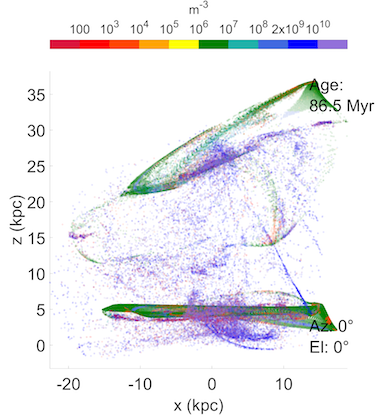}
        }
                \subfigure[]{%
          \includegraphics[width=0.4\textwidth]{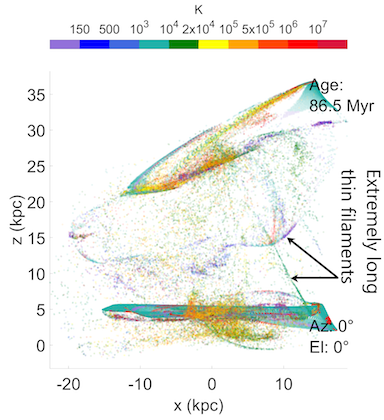}
        }\\ 
                \subfigure[]{%
          \includegraphics[width=0.4\textwidth]{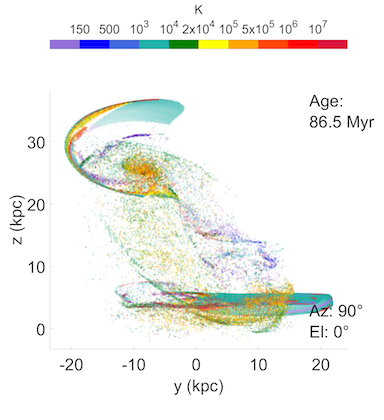}
        }
                \subfigure[]{%
          \includegraphics[width=0.4\textwidth]{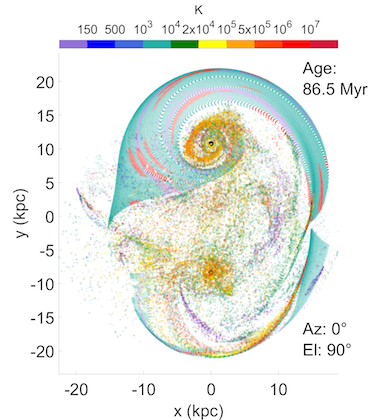}
        }\\ 
    \end{center}
\caption{Late time views of the density (panel a)) and temperature (panels b), c), and d)) in a model with a $20\degree$ inclination and a 10 kpc offset.}%
\label{fig:10kpc20tilt2}
\end{figure*}

\indent As shown in Paper I there are distinct morphological differences between low($<$ few kpc) and high($>$ half galactic radius) impact offset collisions.  It was found that as the disc offset was increased, no longer does the bridge take on an hourglass shape or is a CBD region formed.  Instead, with large offsets the bridge is twisted into a long sheet-like structure.  This is seen again here for the $20 \degree $ inclination, 10 kpc offset, collision shown in \cref{fig:10kpc20tilt1}.  \Cref{fig:10kpc20tilt2} shows this system at an age of 86.5 Myr.   More material remains in the bridge compared to the 500 pc offset collision at this time.  Infall is occurring onto G1, but in this case the region of hot material is less prominent.  The falling material is on a greater arcs around the center of G1, and splashes throughout the disc. I.e., increased disc offsets produce faster infall of material some of which traverses larger orbits about the center of G1.  Several extremely thin and long filaments over 10 kpc in length can be seen in the bridge.

\subsubsection{\texorpdfstring{$45\degree$}{} Inclination Collisions}

\begin{figure*}
     \begin{center}
                \subfigure[]{%
          \includegraphics[width=0.4\textwidth]{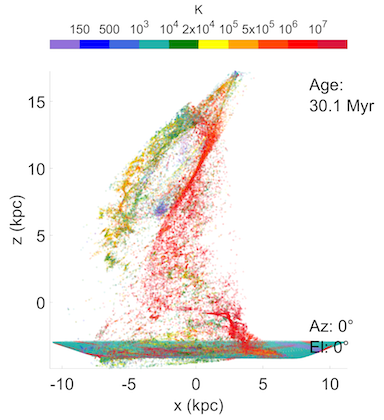}
        }
                \subfigure[]{%
          \includegraphics[width=0.4\textwidth]{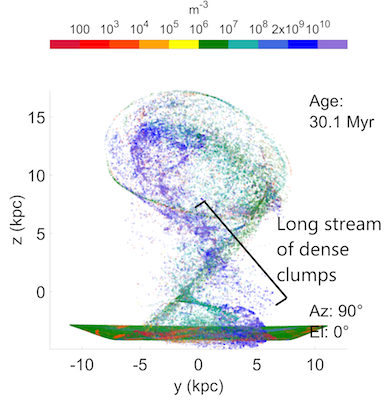}
        }\\ 
                \subfigure[]{%
          \includegraphics[width=0.4\textwidth]{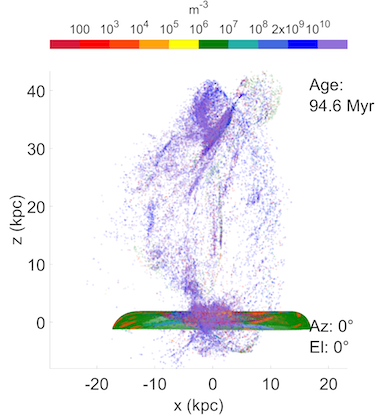}
        }
                \subfigure[]{%
          \includegraphics[width=0.4\textwidth]{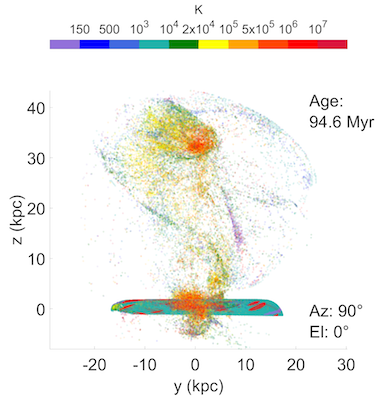}
        }\\ 
    \end{center}
\caption{Edge-on views of the temperature (panels a) and d)) and the density (panels b) and c)) at two times in a model with a $45\degree$ inclination and a 500 pc offset.}%
\label{fig:500pc45tilt1}
\end{figure*}

\indent In disc-disc collisions with an inclination of 45$\degree$, long streams of gas above and below the smaller galaxy G2 begin to form.  Though long filaments of gas also form in low inclination bridges they mostly remain between G1 and G2, whereas the streams in $45\degree$ and greater inclination collisions are pushed out of G2 both above and below its disc.  Also unlike face-on collisions, no CBD-like region forms in the bridge.  This bridge is overall more chaotic than the low inclination collisions, but some layering is still clear as the ISM is separated out by density.  \Cref{fig:500pc45tilt1} shows the 500 pc impact offset $45\degree{}$ inclination collision at 30 Myr and at 94.6 Myr.  Frame b) highlights a region of dense clumps, streaming from G2 into the bridge.  The location of a large dense region is similar to the prominent dense region of gas in the top of the Taffy Galaxies bridge (though in the lower galaxy here).  The $20\degree{}$ and $45\degree{}$ runs both show similarities to the Taffy Bridge at around 30 Myr, depending on which line of sight is chosen to view the bridges.  The bridges formed from inclinations beyond $45\degree{}$ do not resemble the Taffy, reaffirming previous constraints on the inclination of the Taffy collision as less than $45\degree$. 

\indent For all collisions with inclinations of less then $20 \degree$, the stellar discs are mostly separate from the splash bridges.  In Appendix A the stellar discs for each run are shown for comparison.  For the $45\degree, 65\degree$ and $90\degree$ inclination runs the gravitational interaction is significant enough to remove stars from G2 and spread them across parts of the bridge.

\begin{figure*}
     \begin{center}
                \subfigure[]{%
          \includegraphics[width=0.4\textwidth]{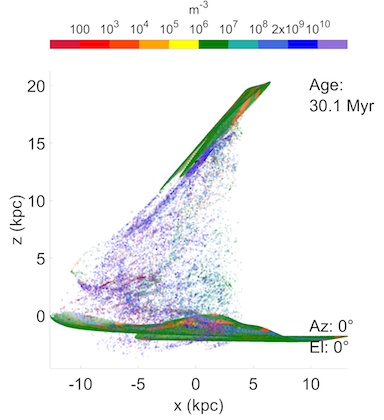}
        }
                \subfigure[]{%
          \includegraphics[width=0.4\textwidth]{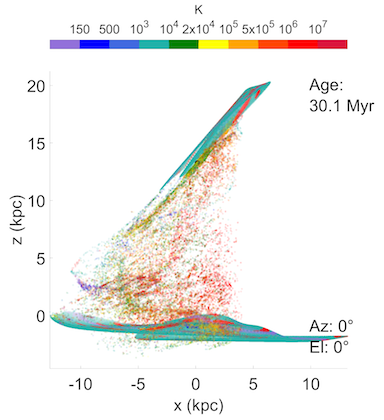}
        }\\ 
                \subfigure[]{%
          \includegraphics[width=0.4\textwidth]{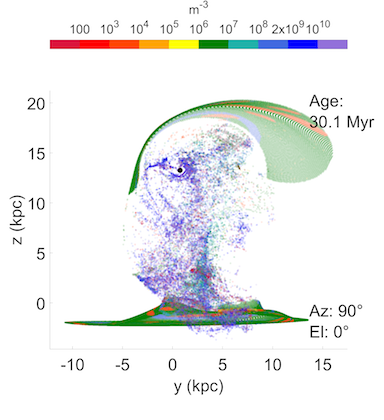}
        }
                \subfigure[]{%
          \includegraphics[width=0.4\textwidth]{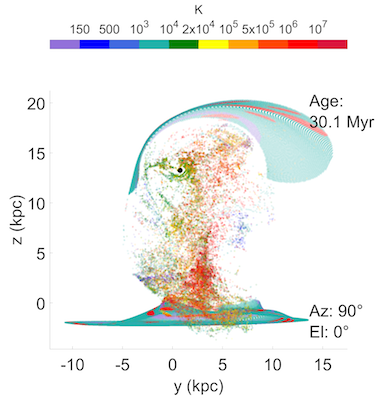}
        }\\ 
    \end{center}
\caption{Edge-on views of the density (panels a) and c)) and the temperature (panels b) and d)) in a model with a $45\degree$ inclination and s 10 kpc offset. All at a time of 30.1 Myr.}%
\label{fig:10kpc45tilt}
\end{figure*}

\subsubsection{\texorpdfstring{$65\degree$}{} Inclination Collisions}

\begin{figure*}
     \begin{center}
                \subfigure[]{%
          \includegraphics[width=0.4\textwidth]{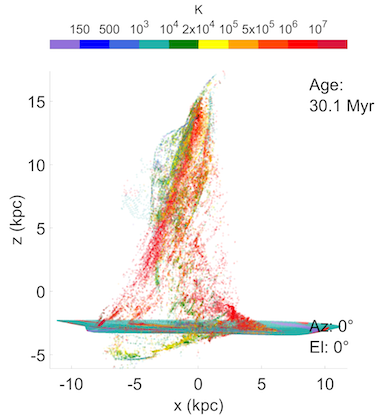}
        }
                \subfigure[]{%
          \includegraphics[width=0.4\textwidth]{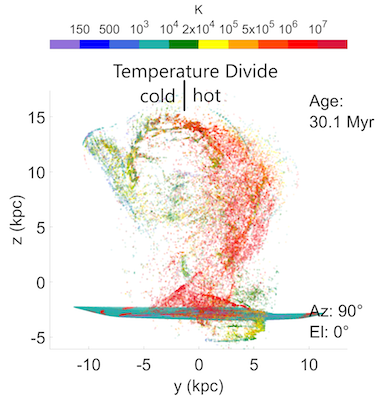}
        }\\ 
                \subfigure[]{%
          \includegraphics[width=0.4\textwidth]{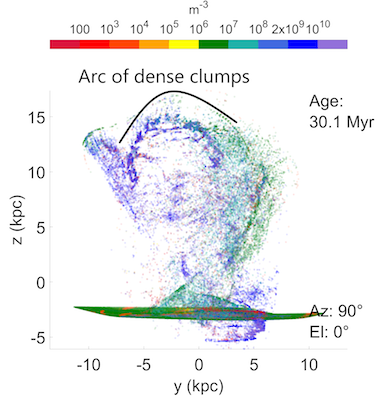}
        }
                \subfigure[]{%
          \includegraphics[width=0.4\textwidth]{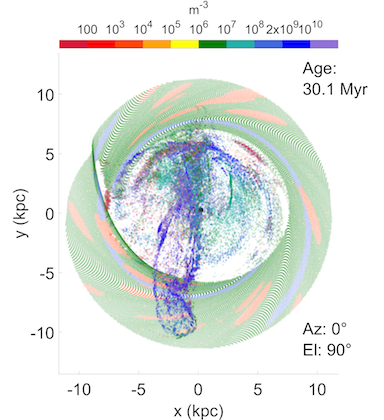}
        }\\ 
    \end{center}
\caption{Edge-on (panels a)-c)) and face-on (panel d)) views of the temperature (top row) and density (bottom row) of a model with a $65\degree$ inclination and a 500 pc offset.}%
\label{fig:500pc65tilt}
\end{figure*}

\begin{figure*}
                \subfigure[]{%
          \includegraphics[width=0.4\textwidth]{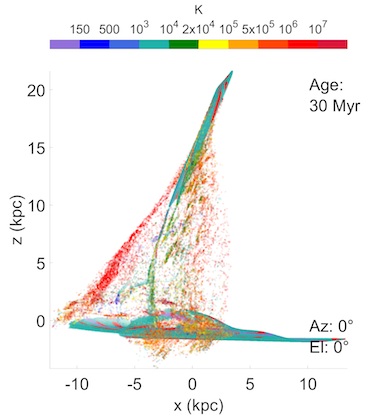}
        }
                \subfigure[]{%
          \includegraphics[width=0.4\textwidth]{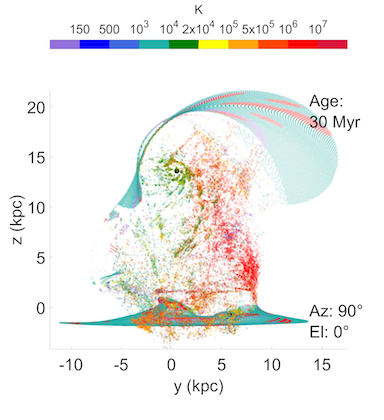}
        }\\ 
                \subfigure[]{%
          \includegraphics[width=0.4\textwidth]{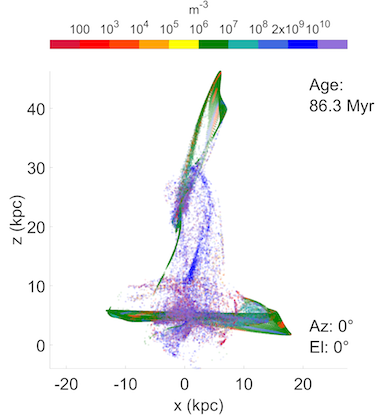}
        }
                \subfigure[]{%
          \includegraphics[width=0.4\textwidth]{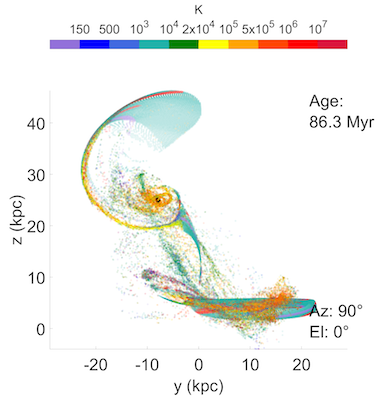}
        }\\ 
  \caption{Edge-on views of the temperature (panels a, b) and d)) and density (panel c)) at two times in a model with a $65\degree$ inclination and a 10 kpc offset.}%
  \label{fig:10kpc65tilt1}
\end{figure*}

\indent In the $65\degree$ inclination collisions shown in \cref{fig:500pc65tilt,fig:10kpc65tilt1} streams are created on each side of the gas disc of G2 as G2 passes through G1.  The stellar disc for this run at 30 Myr can be seen in \cref{fig:10kpcstellar} and lines up with much of left side of bridge seen in \cref{fig:10kpc65tilt1} (b).  The stars in this run appear closer to much of the gas in the bridge region than in the lower inclination collisions though there is still a few kpc displacement between the two components along the x-axis.  

\indent \Cref{fig:500pc65tilt} c) shows the density of the gas at 30 Myr after the discs have collided.  This image shows a long arc of dense clouds in purple across the top of the image.  On the right of the image there are long filaments of \SI{e10} \dunit{} gas surrounded in the light blue by gas of density \SI{e7} \dunit.  \Cref{fig:10kpc65tilt1} shows the collision with a 10 kpc offset, and frames c) and d) show the system at a much later age of 86 Myr.  In b) we can see the divide in temperature of the gas clouds.  Along the left of the bridge are high density low temperature clouds in light blue, and on the right of the bridge the clouds remain millions of degrees K with lower \SI{e7} \dunit densities.  The gas in the bridge at the late time has either accreted onto G1, or remains in long 20 kpc long filaments stretched from G2 down to G1.  Frame d) shows these long filaments, which are similar to one seen in the Arp 194 (UGC 6945) system.  In Arp 194 a long stream of star forming clumps appears to be stretched from one galaxy down toward a spiral below it.  

\subsubsection{\texorpdfstring{$90\degree$}{} Inclination Collisions}
\label{sec:90degrees}

\begin{figure*}
     \begin{center}
                \subfigure[Face on view of G1.]{%
          \includegraphics[width=0.4\textwidth]{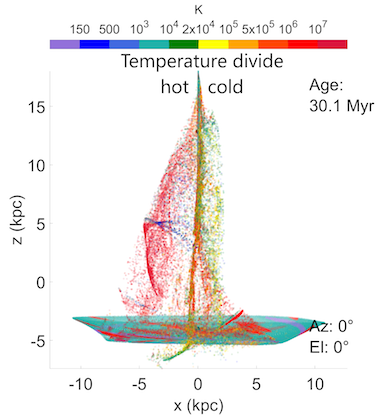}
        }
                \subfigure[Face on view of G2.]{%
          \includegraphics[width=0.4\textwidth]{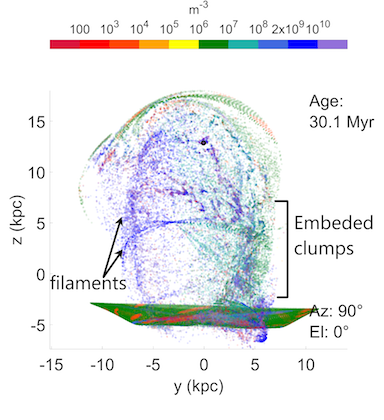}
        }\\ 
                \subfigure[]{%
          \includegraphics[width=0.4\textwidth]{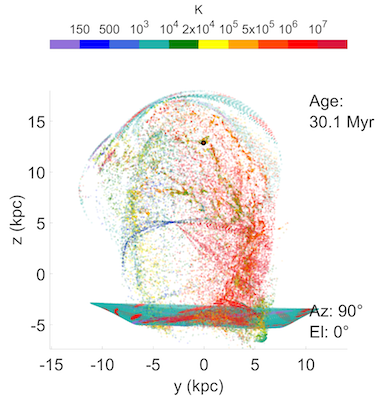}
        }
                \subfigure[]{%
          \includegraphics[width=0.4\textwidth]{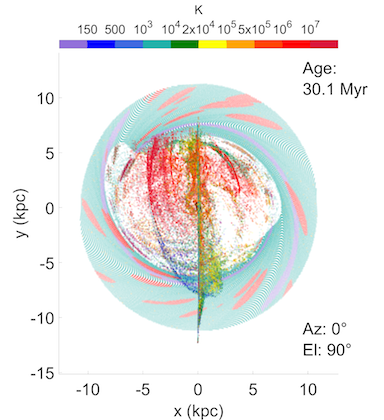}
        }\\ 
    \end{center}
\caption{Edge-on (panels a)-c)) and face-on (panel d)) views of the temperature (panels a), c), d)) and density (panel b)) of a model with a $90\degree$ inclination and a 500 pc offset.}%
\label{fig:500pc90tilt1}
\end{figure*}

\indent \Cref{fig:500pc90tilt1} shows the $90\degree{}$ collision at 30 Myr.  Gas streams develop as G2 cuts through the centre of G1.  The reason gas is pushed more into the plane of G1 is because G1 has a somewhat higher gas surface density and mass in its disc.  Had G2 been the dominant galaxy in this collision gas would be pushed more toward its plane of rotation.  Each stream of gas is separated by its initial gas phase which is not visible in the figures.  Much of the gas from several initial density components reaches a final density of $\SI{2e9} \dunit$ as it cools.  

\indent In \Cref{fig:500pc90tilt1} a) we can see a clear divide in gas temperatures on each side of G2, this time above and below the disc of G2.  Gas of G2 that impacts the central region of G1 experiences stronger shocks as well as more subsequent collisions,  where the gas on the right of G2 in frame a) passed through the outer disc of G1 and was not shocked as strongly.  This allowed the gas right of the G2 disc to cool whereas the gas on the left of G2 is still hot.  Frame b) shows gas densities.  We can see the gas which has already cooled to 20,000 K, has also condensed into filaments with densities of \SI{e10} \dunit.  The stream gas to the right in frame b) is of order \SI{e7} \dunit{}, and is not found within tight filaments.  Large (kpc) size clumps of dense \SI{e10} \dunit gas can be seen within the stream.  In frame c) and d) the two temperature populations of 20,000 K and \SI{e6} K gas can be seen as well.  The rotation of G2 in frame c) is out of the page, again showing that cooling in the bridge seems to be helped by a rarefaction in the gas cloud ensemble.  

\indent The stellar disc of G2 for this collision is gravitationally stretched out toward G1, though it does not follow the gas streams.

\begin{figure*}
     \begin{center}
                \subfigure[]{%
          \includegraphics[width=0.4\textwidth]{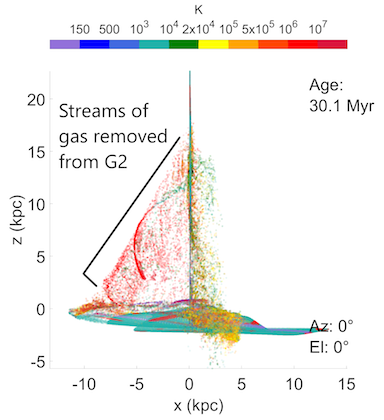}
        }
                \subfigure[]{%
          \includegraphics[width=0.4\textwidth]{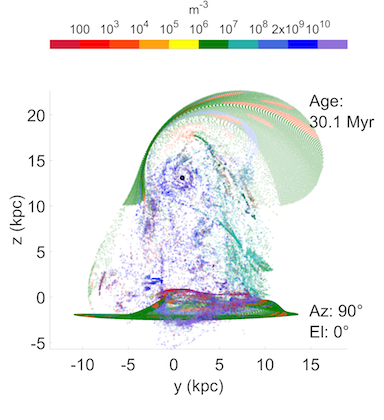}
        }\\ 
                \subfigure[]{%
          \includegraphics[width=0.4\textwidth]{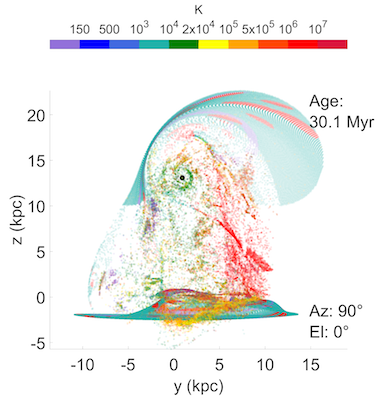}
        }
                \subfigure[]{%
          \includegraphics[width=0.4\textwidth]{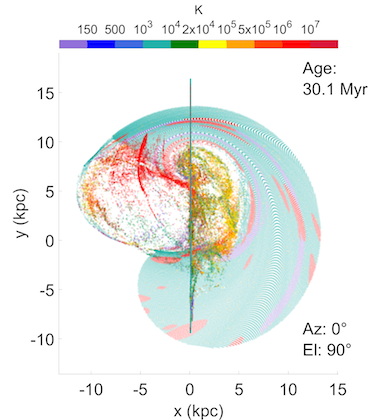}
        }\\ 
    \end{center}
\caption{Edge-on (panels a)-c)) and face-on (panel d)) views of the temperature (panels a), c), d)) and density (panel b)) of a model with a $90\degree$ inclination and a 10 kpc offset.}%
\label{fig:10kpc90tilt1}
\end{figure*}

\indent \Cref{fig:10kpc90tilt1} (a) shows more pronounced gas streams than in the 500 pc impact offset $90\degree$ inclination collision.  Here the 10 kpc offset leads to the outer disc of galaxy G2 passing near the centre of G1.  Thus, the gas and star particles of G2 are more loosely bound when they pass through a much stronger part of the G1 gravitational potential, causing much greater disturbance.  The stars are not thrown above or below the disc of either galaxy, though the stellar disc of G2 is tidally stretched. A substantial amount of gas on the other hand is thrown far from the disc of G1 and G2.  The asymmetry of gas streams in \cref{fig:10kpc90tilt1} is because of the 10 kpc offset.  Most of the collision occurs on one side of the galaxy, so gas is only thrown in one direction.  In the 500 pc offset case gas was thrown in both directions much more symmetrically.  

\begin{figure*}
     \begin{center}
                \subfigure[]{%
          \includegraphics[width=0.4\textwidth]{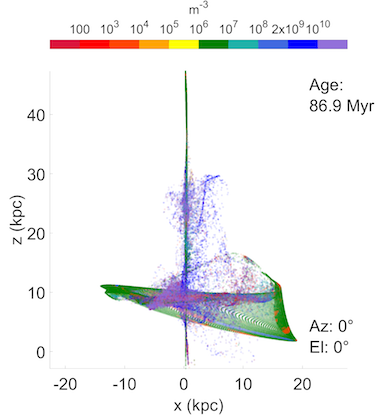}
        }
                \subfigure[]{%
          \includegraphics[width=0.4\textwidth]{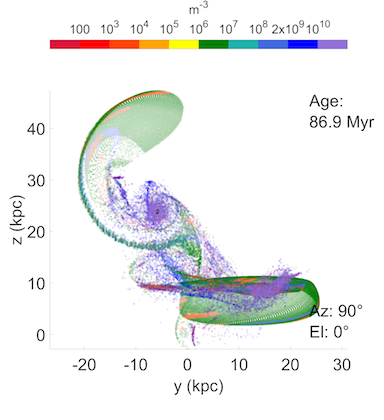}
        }\\ 
                \subfigure[]{%
          \includegraphics[width=0.4\textwidth]{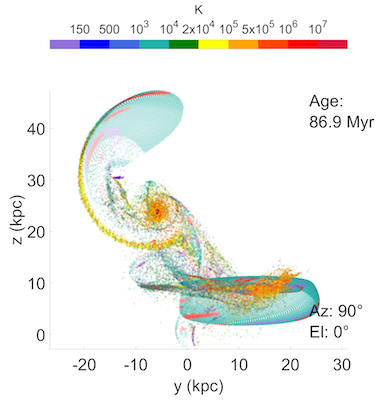}
        }
                \subfigure[]{%
          \includegraphics[width=0.4\textwidth]{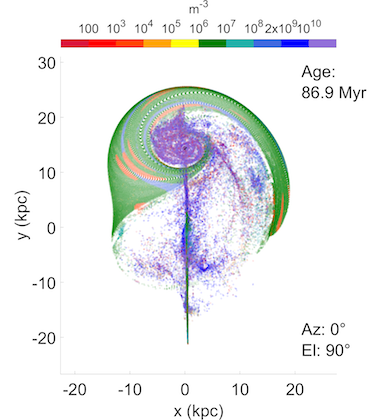}
        }\\ 
    \end{center}
\caption{Edge-on (panels a)-c)) and face-on (panel d)) views of the density (panels a), b), d)) and density (panel c)) of a model with a $90\degree$ inclination and a 10 kpc offset, at a late time.}%
\label{fig:10kpc90tilt2}
\end{figure*}

\subsubsection{\texorpdfstring{$90\degree$}{} Inclination Collision - with G2 initialized in the x-z plane}

\begin{figure*}
     \begin{center}
                \subfigure[]{%
          \includegraphics[width=0.4\textwidth]{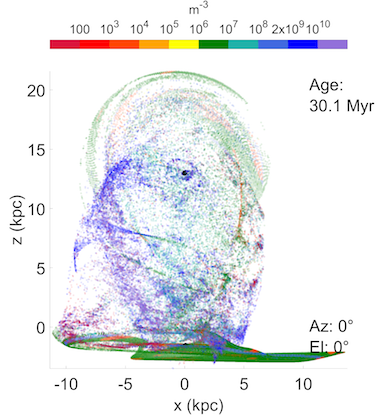}
        }
                \subfigure[]{%
          \includegraphics[width=0.4\textwidth]{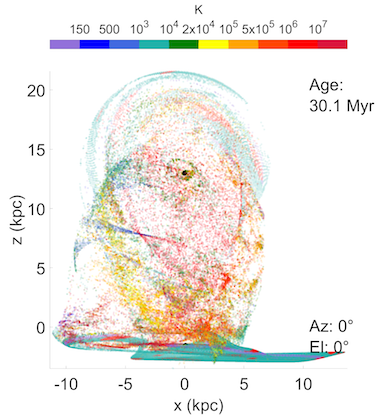}
        }\\ 
                \subfigure[]{%
          \includegraphics[width=0.4\textwidth]{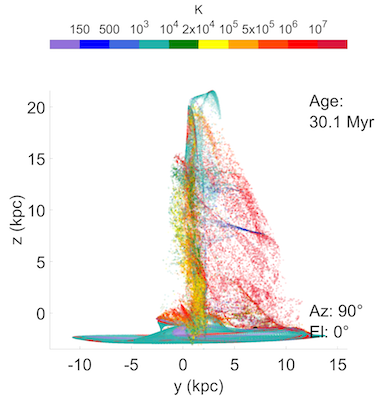}
        }
                \subfigure[]{%
          \includegraphics[width=0.4\textwidth]{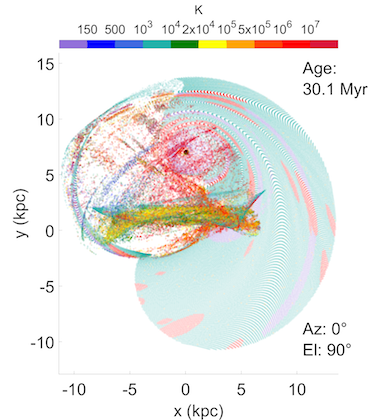}
        }\\ 
    \end{center}
\caption{Views of the density (panel a) and temperature distributions for a model with a $90\degree$ inclination and a 10 kpc offset.  In this case G2 was initialized in the x-z plane, rather than the y-z of previous runs.}%
\label{fig:10kpc90tiltrotated}
\end{figure*}

\begin{figure*}
     \begin{center}
                \subfigure[]{%
          \includegraphics[width=0.4\textwidth]{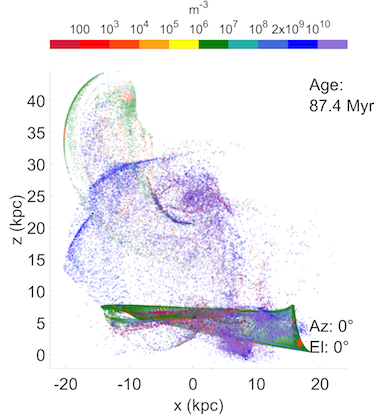}
        }
                \subfigure[]{%
          \includegraphics[width=0.4\textwidth]{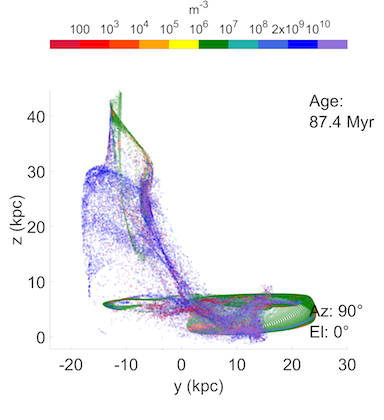}
        }\\ 
                \subfigure[]{%
          \includegraphics[width=0.4\textwidth]{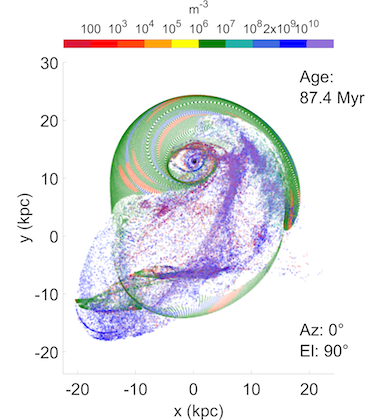}
        }
                \subfigure[]{%
          \includegraphics[width=0.4\textwidth]{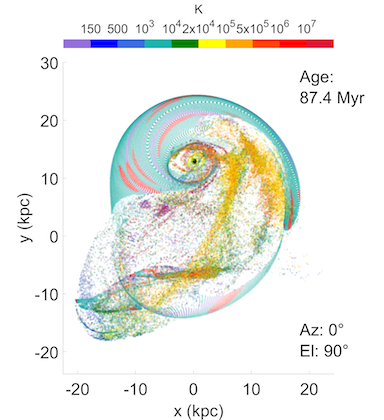}
        }\\ 
    \end{center}
    \caption{Late time views of the density (panels a)-c)) and temperature distributions in a model with a $90\degree$ inclination and a 10 kpc offset. G2 was initialized in the x-z plane.}%
     \label{fig:10kpc90tiltrotated2}
\end{figure*}

\indent By rotating the disc of G2 90$\degree{}$ about the z-axis relative to the models of the previous section, none of the gas in G2 will pass near the dense center of G1.  This results in a slight weakening of the fastest shocks in G2 but a strengthening of the shocks to G1 gas.  Many clumps and filamentary structures are produced in this bridge and can be seen in \cref{fig:10kpc90tiltrotated} at 30 Myr.  A higher fraction of the gas in G2 is shocked than compared to the other 90$\degree{}$ inclination collisions.  \Cref{fig:10kpc90tiltrotated2} d) shows this system with a significant amount of gas of temperature \SI{e5}-\SI{e6} K in the splash bridge between G1 and G2 remaining at 87.4 Myr, but less of temperature > \SI{e6}K than compared to the previous $90\degree{}$ collision.  

\FloatBarrier
\clearpage
\subsection{Shock and Mach Number Distributions}
\label{sec:distributions}

\begin{figure}
        \subfigure[]{%
           \includegraphics[width=0.4\textwidth]{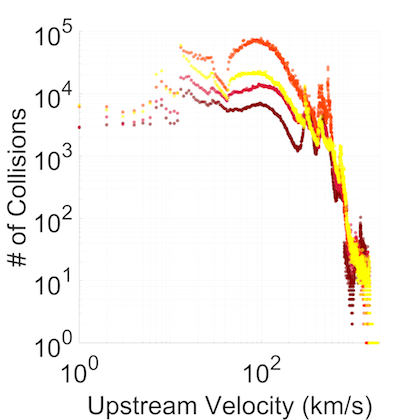}
        }\\ 
        
        \subfigure[]{%
           \includegraphics[width=0.4\textwidth]{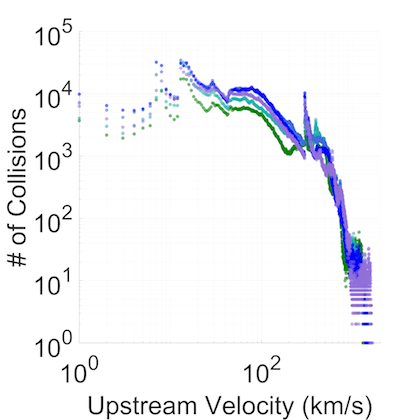}
        }\\ 
\caption{Total count of cloud collisions over the entire 150 Myr run time, with velocity bin size of 1 \kms.  a) 500 pc impact offset runs with inclinations as follows, dark red 20$\degree$, red 45$\degree$, orange 65$\degree$, yellow 90$\degree$.  b) 10 kpc offset runs with inclinations: green 20$\degree$, light turquoise 45$\degree$, light blue 65$\degree$, dark blue 90$\degree$, purple 90$\degree${} inclination with G2 rotated 90$\degree${} on y-axis.}%
\label{fig:totaldists}
\end{figure}

\begin{figure}
   \includegraphics[width=0.4\textwidth]{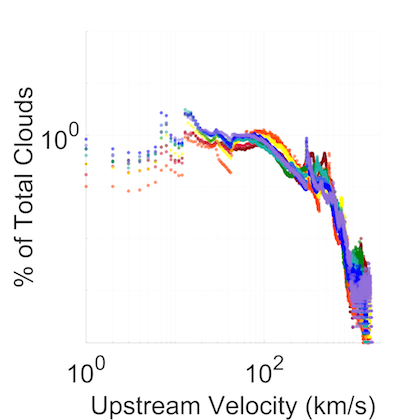}
\caption{Counts of cloud collisions normalized to the total number of collisions that occur in the run over the entire 150 Myr run time, the velocity bin size is 1 \kms.  500 pc offsets with inclinations: dark red 20$\degree$, red 45$\degree$, orange 65$\degree$, yellow 90$\degree$, 10 kpc offsets with inclinations, green 20$\degree$, light turquoise 45$\degree$, light blue 65$\degree$, dark blue 90$\degree$, purple 90$\degree${} inclination with G2 rotated 90$\degree${} about the z-axis.}%
\label{fig:fracdists}
\end{figure}

\indent \Cref{fig:totaldists} shows the counts of cloud-cloud collisions in velocity bins, over all the runs, and integrated over the entire 150 Myr run times.  The lowest number of collisions occurs in the 20\degree{} inclination runs for both 500 pc and 10 kpc impact offsets.  As the inclination is increased, so too is the total number of collisions.  The trend of increasing number of collisions is consistent with expectations based on such simple geometric considerations.  When the rotation planes of each galaxy are no longer similar, gas cloud orbits end up in a much more chaotic distribution increasing the likelihood of cloud-cloud collisions.  We also expect that the low inclination runs should produce a heightened number of shocks with velocities of several hundred \kms, since the counter-rotating discs leads to larger shock velocities.  \Cref{fig:totaldists} a) shows that the 500 pc impact offset, $20\degree{}$ inclination run does produce the most shocks in the several hundred \kms range.  The 10 kpc offset runs in frame b) do not show a heightened number of 500 \kms shocks. However, \Cref{fig:fracdists} does show the 10 kpc $20\degree{}$ run produces some of the highest fractions of shocks with velocities greater than 100 \kms.

\begin{figure*}
     \begin{center}
                \subfigure[]{%
          \includegraphics[width=0.4\textwidth]{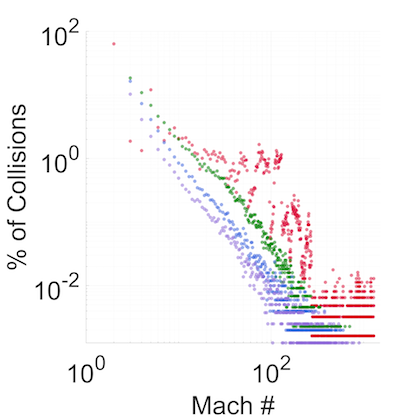}
        }
                \subfigure[]{%
          \includegraphics[width=0.4\textwidth]{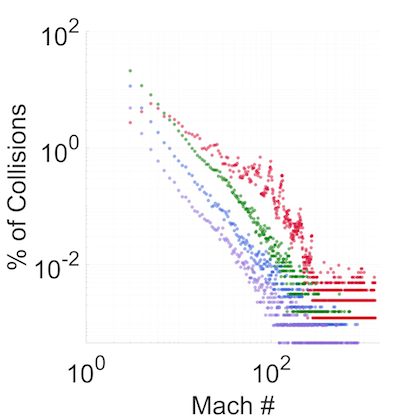}
        }\\ 
                \subfigure[]{%
          \includegraphics[width=0.4\textwidth]{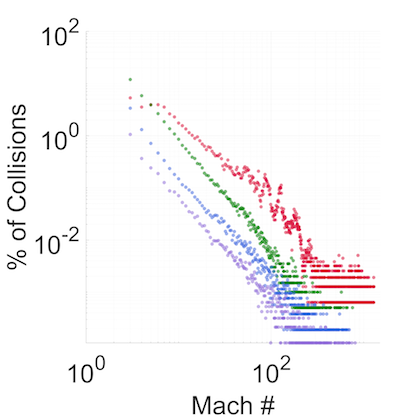}
        }
                \subfigure[]{%
          \includegraphics[width=0.4\textwidth]{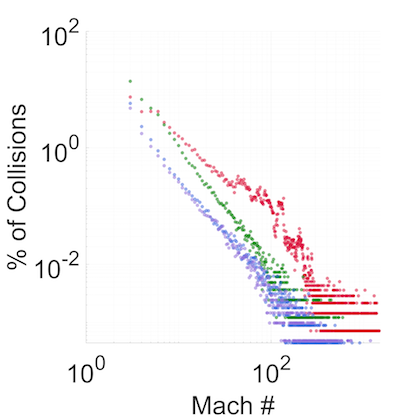}
        }\\ 
    \end{center}
    \caption{Each point is the number of shocks in a given Mach number bin, with bins of width one unit. Color indicates collisions that occur within a 10 Myr period.  Red 20-30 Myr, green 60-70 Myr, blue 100-110 Myr and purple 140-150 Myr.  All frames are 500 pc offset runs with a) 20$\degree$, b) 45$\degree$, c) 65$\degree$, and d) 90$\degree${} inclinations.  The Mach numbers are determined by the speeds at which gas flows across the shock wave in the frame of the shock.}%
     \label{fig:earlyandlatedistmachs}
\end{figure*}

\begin{figure*}
     \begin{center}
                \subfigure[]{%
          \includegraphics[width=0.4\textwidth]{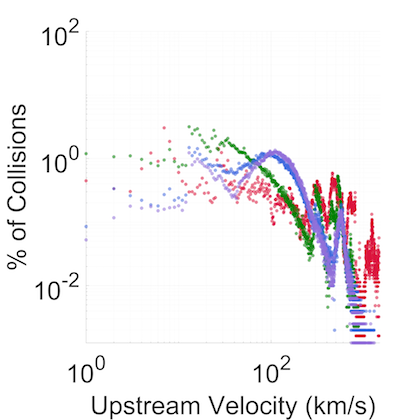}
        }
                \subfigure[]{%
          \includegraphics[width=0.4\textwidth]{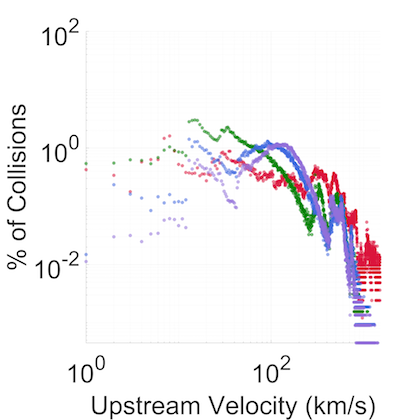}
        }\\ 
                \subfigure[]{%
          \includegraphics[width=0.4\textwidth]{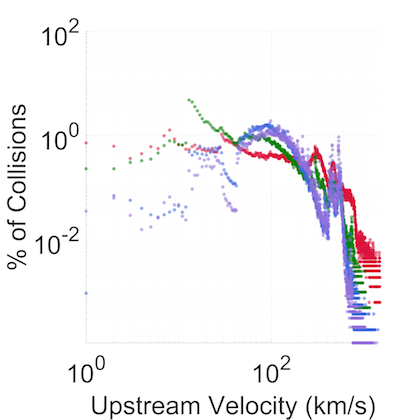}
        }
                \subfigure[]{%
          \includegraphics[width=0.4\textwidth]{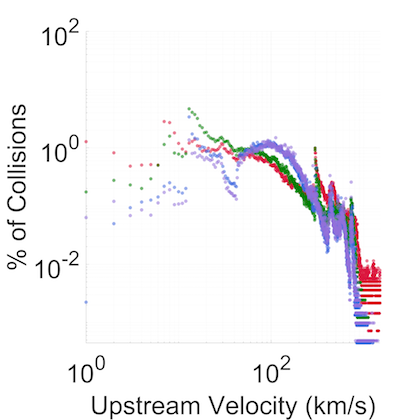}
        }\\ 
    \end{center}
    \caption{Each point is the number of shocks in a given upstream velocity bin, with bins of width 1 \kms. Color indicates collisions that occur within a 10 Myr period.  Red 20-30 Myr, green 60-70 Myr, blue 100-110 Myr and purple 140-150 Myr.  All frames are 500 pc offset runs with a) 20$\degree$, b) 45$\degree$, c) 65$\degree$, and d) 90$\degree${} inclinations.  The upstream velocities are determined by the speeds at which gas flows across the shock wave in the frame of the shock.}%
     \label{fig:earlyandlatedistshocks}
\end{figure*}

\indent We find a surprising result in the distributions of upstream (pre-shock) velocity of colliding clouds - each run follows a similar form when normalized to the total number of collisions that occurred.  \Cref{fig:fracdists} shows the normalized distributions of nine runs stacked together.  This implies, independent of collision offset and inclination, that for each high velocity shock there are a proportional number of lower velocity shocks.  \Cref{fig:earlyandlatedistmachs} shows the distribution of shock Mach numbers within colliding clouds that occur within 10 Myr time intervals for 500 pc offset runs of various inclinations.  This reveals that the total distribution is not created from the main collision, rather the form takes shape after some time (~20 Myr) has passed and turbulence can build up the lower end of the graph.  This also shows us that every run produces continued turbulence for the entire 150 Myr run time.  The red curves show collisions which happen during the direct collision between the galaxy discs.  The green, blue and purple curves are 10 Myr time bins: 60-70, 100-110, 140-150 Myr, respectively. The lower the inclination the more unformed the red time bin becomes.  \Cref{fig:earlyandlatedistshocks}, shows the distribution of upstream shock velocities over time, measured in the frame of the shocks.  Heightened counts of upstream velocities greater than 100 \kms are indicative of large scale collision events, such as the initial collision between the two galactic gas discs.  For the higher inclinations there is no singular global collision event across the galaxies since it now takes nearly 10 Myr for the disc of G2 to pass entirely through G1.  This results in a smoother distribution of upstream velocities in the red time bin as the inclination is increased.  We can also see a quicker build up in the lower end of the shock distribution as inclination is increased.  The $90\degree{}$ case seems to lock in nearly immediately.

\indent Seen in red in \cref{fig:earlyandlatedistmachs}, 20-30 Myr into the integration when the galaxies make their nearest approaches there is clearly a heightened number of shocks at the highest velocities.  The lower $20\degree{}$ inclination shows the highest fraction of shocks in this 20-30 Myr time bin.  This distribution weakens slightly with time in every run, but there is still a substantial number of collisions occurring 120 Myr after the disc on disc collisions.

\indent \citet{almataffy2019} analyzed shock distributions in ALMA observations of the Taffy Galaxies and found that, \textit{"a significant amount of the high-velocity component of the bridge gas was consistent with shocks with velocities of 200-300 km/s."} Frame a of \cref{fig:earlyandlatedistmachs}, in red, shows the 10-20 Myr time bin after the gas discs have collided. Shown in \cref{fig:earlyandlatedistshocks} are significantly heightened shock counts between 200-300 \kms.  In the same ALMA data the molecular material was found to be highly disturbed; clumps in the filaments contained broad line widths of 80-150 \kms.  When separating the shocks by ISM phase (here density and temperature), we see that during the same 10-20 Myr post-collision time bin there is a peak in the counts near 100 \kms{} for the \hh{} gas.

\begin{figure*}
     \begin{center}
                \subfigure[]{%
          \includegraphics[width=0.4\textwidth]{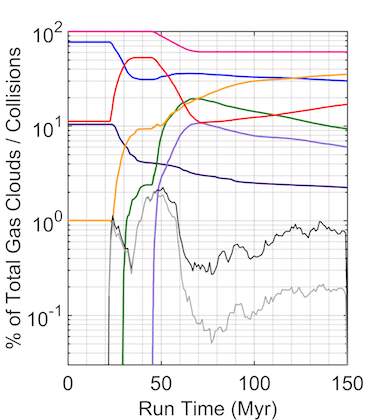}
        }
                \subfigure[]{%
          \includegraphics[width=0.4\textwidth]{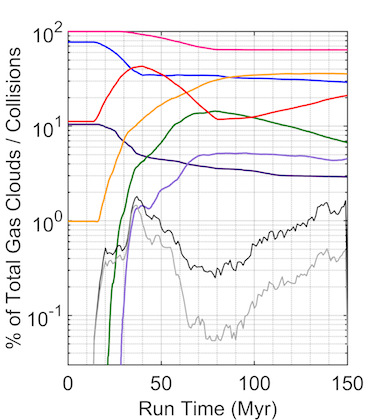}
        }\\ 
                \subfigure[]{%
          \includegraphics[width=0.4\textwidth]{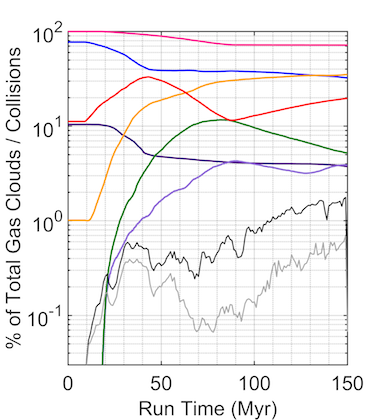}
        }
                \subfigure[]{%
          \includegraphics[width=0.4\textwidth]{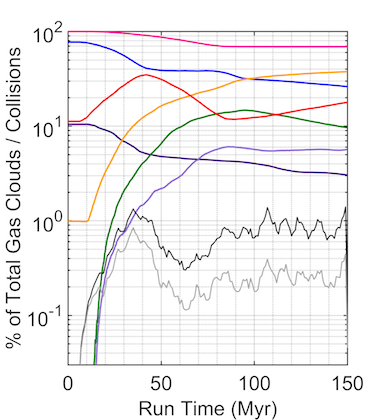}
        }\\ 
    \end{center}
    \caption{Cloud populations over time for 500 pc impact runs at different inclinations.  Frame a) is 20$\degree${} b) 45$\degree${} c) 65$\degree${} d) 90$\degree${} inclination.  Colors: dark purple represents gas clouds 50-150 K (\hh{} and CNM components initially in this bin), light purple 150-\SI{e3}K, blue \SI{e3}-\SI{e4} K (\HI{} initially here), green \SI{e4}-\SI{2e4} K, orange \SI{2e4}-\SI{e6} K (WNM initially here), red above \SI{e6} K (hot diffuse initially), pink fraction of gas clouds where dust grains survive, black is the percent of total shocks and grey is the percent of shocks above 250 \kms{} in each 1 Myr time bin.}%
     \label{fig:cloudpopulations}
\end{figure*}

\indent Cloud temperature distributions and shock counts over time for the 500 pc impact offset runs are shown in \cref{fig:cloudpopulations}.  The collision rate, in black, shows a quick rise to a peak, which occurs as the gas discs pass through one another.  Only in the $65\degree{}$ inclination is there no semblance of a peak in the collision rate.  Instead, there is a rapid increase to a constant rate for about 50 Myr before a return to a slowly increasing collision rate.  After the discs make their initial impact the number of collisions per time falls to a reduced rate.  In grey the number of collisions that exceed 250 \kms{} can be seen to closely track the total collision rate after the disc on disc collision, which is consistent with the Mach distributions shown in \cref{fig:earlyandlatedistmachs}.

\indent In the light purple curve we see the fraction of gas clouds with temperatures between 150 and 1,000 K, grows extremely quickly.  None of the gas in the galaxies is initially in this temperature bin, so this large fraction of gas clouds can only come from cloud collisions that have cooled.  The fraction of the hottest gas, $T>10^6 K$, in the red curve, quickly increases to a maximum gas fraction during the disc on disc collision.  The hot gas fraction then decreases to the initial fraction after 50-70 Myr, before increasing again at a slower rate from bridge turbulence.  In the lower inclination collisions where shock velocities are slightly elevated, we see that the fraction of the hot gas phase decreases more slowly.  Interestingly, the temperature bin shown by the orange curve, $20,000<T<10^6 K$, is never decreasing.  Changes in the orange temperature bin precedes the green bin by about 5 Myr and the light purple bin by about 25 Myr.  The delay between the orange, green, and light purple bin shows the approximate average cooling times between each bin of temperature.  Many of the gas clouds are in the green bin at the end of the run, with $10,000<T<20,000 K$.  These are mostly gas clouds that have stopped net cooling, with the UV heating balancing the cooling.  When this occurs the only way for a cloud to change its temperature is through heating in further collisions.  

\indent The pink line near the top of each graph shows the fraction of clouds that contain surviving dust particles.  Survival here is determined by whether or not they have sputtered below a grain radius of 1 nm.  Across all runs, much of the dust survives.  The lowest fraction is $65\%$ in the 500 pc offset $20\degree$ inclination collision.

\flushbottom
\clearpage

\section{Conclusion}

Splash bridges are chaotic messy environments. We have developed an efficient sticky particle simulation code with a variety of heating and cooling processes included to model their early evolution and compare to observation. Generally, we find that inclination and impact offset effects leave behind traceable global morphologies within chaotic bridges.  Alternately, the observable kinematics and turbulence within the bridges may be used to loosely trace back to the inclination and offset impact parameters up to hundreds of Myr after the initial impact. For example, a collision between two outer discs leaves behind a distinctly different splash bridge seen from most line of sights than a low offset collision. In the remainder of this section we summarize more specific results.

\subsubsection{Bridge morphology}
\begin{enumerate}
\item The low inclination collisions ($<20\degree$) described above display some similar bridge structures in the first 20 Myr of evolution to those modeled in Paper I, where we used a simpler version of the simulation code.  Later, the increased turbulence modeled here causes significant deviations in splash bridge evolution.  

\item Low inclination collisions with small impact offsets form a central bridge disc (CBD), a large flat region of gas near the centre of mass of the two galaxies.  Long twisted filaments fill the rest of the splash bridge between the galaxies.

\item Low inclination collisions with large impact offsets produce long twisted sheets along with filaments of gas between the two galaxies.

\item For collisions with inclinations of $45\degree$ to $90\degree$, rarefaction and compression of large regions of gas occurs after the discs collide.  The centers of each galaxy pass one another at around 550 \kms.  This is much greater than the maximum rotational velocity of either galaxy which is between 200-300\kms.  The rotation of G2 strengthens or weakens the initial collision shocks and shapes the type of turbulence to follow.  Gas clouds in G2 rotating with a positive z-direction component (parallel to the net G2 z-velocity), crash into G1 as the discs collide, and initiate a pile-up in gas clouds from the two discs which causes many more collisions to occur after the main disc-disc collision. This prevents these gas clouds from cooling. On the other hand, gas clouds from G2 with negative z-velocities, impact at lower velocities, and are subsequently rarefied when they fall back to G1.  The increasing distance between gas clouds prevents continued turbulence which allows these gas clouds to cool.

\item Gravitational tidal effects on the stellar discs produce stronger perturbations in the high inclination collisions and are strongest in the 10 kpc offset inclined collisions (due to stronger coupling with disc rotation).  This leads to the stellar component appearing to blend with the gas bridges in the 10 kpc offset runs 45$\degree{}$, 65$\degree{}$, and $90\degree{}$ inclination collisions along some lines of sight.  However, the gas bridge is still displaced from the stellar disc by varying amounts.  In no case does the stellar disc fully overlap the splash bridge.  

\subsubsection{Shock distributions and ISM populations}
\item The distributions of cloud collision Mach numbers in colliding clouds relax to similar profile forms in all nine runs. Specifically, independent of the impact parameter, each run when normalized to the total number of collisions follows a very similar form for shocks with velocity jumps ranging from 10 \kms{} to 200 \kms.  Above 200 \kms{} the spread between distributions in different runs appears to grow.

\item The maximum shock velocities get smaller as the inclination between gas discs is increased.  But the overall number of collisions per time increases over a time 20 Myr after the gas discs have passed one another.

\item The fraction of hot gas to total gas in the splash bridges decreases over time in all models after the disc on disc collision.  However from 0-20 Myr after the gas discs have collided turbulence drives the hot gas fraction component up.  At the end of each 150 Myr integration time the fraction of hot gas is still increasing.

\item The total number of collisions increases as inclination increases.  This is due to the rotation planes of each galaxy causing extra mixing of momentum in all directions in the splash bridge.  This heightened kinematic turbulence can be seen to increase the fraction of gas in the permitted line cooling regime $(T>20,000K)$.  The strengthening turbulence in higher inclination collisions lowers the overall \HI{} fraction.

\item Recent ALMA observations \citep{almataffy2019} have resolved filamentary structure and clumping of the \HI{} gas across the bridge of the Taffy Galaxies.  The inelastic collisions of the multi-phase gas reproduces bridges with similar filament distributions.

\item In the time period 10-20 Myr after the gas discs have collided, we see a large portion of upstream shock velocities above 200 \kms{} for low inclination cases.  This is consistent with the observed emission of \citet{almataffy2019}, which indicates a high velocity component of 200-300 \kms in the Taffy galaxies.

\item The high density of the \hh{} component in our initial discs results in lower shock velocities than 200-300 \kms, when such clouds are hit by a lower density component at high velocity.  As a result, most molecular material that collides experiences upstream shock velocities consistent with the 80-150 \kms observed in the Taffy.

\end{enumerate}

\indent Given the overall qualitative agreement with the well-studied Taffy system, we expect that these models will be very useful in studying other systems with direct collisions between gas-rich discs. We further hope they will stimulate the acquisition of new, multi-wavelength observations of such systems.


\nocite{hernquist1990}
\nocite{smith15}
\nocite{wang15}
\bibliography{travslibrary}{}
\bibliographystyle{mnras}

\appendix
\section{{Stellar Discs}}

\begin{figure*}
     \begin{center}
                \subfigure[]{%
          \includegraphics[width=0.4\textwidth]{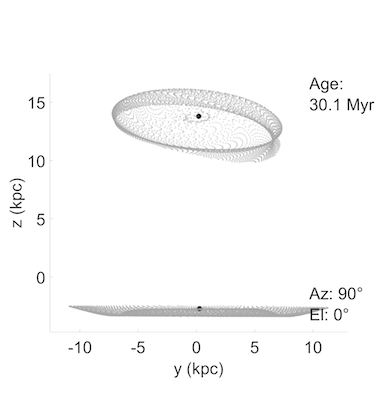}
        }
                \subfigure[]{%
          \includegraphics[width=0.4\textwidth]{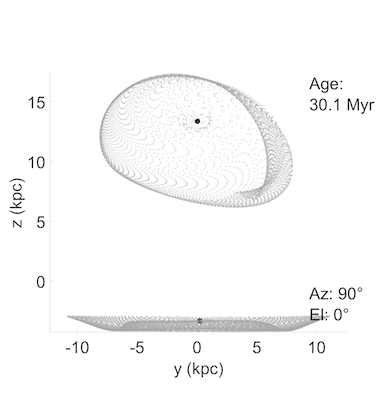}
        }\\ 
                \subfigure[]{%
          \includegraphics[width=0.4\textwidth]{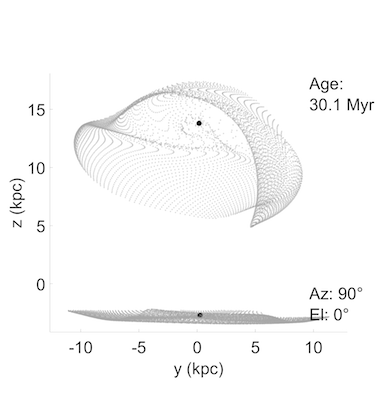}
        }
                \subfigure[]{%
          \includegraphics[width=0.4\textwidth]{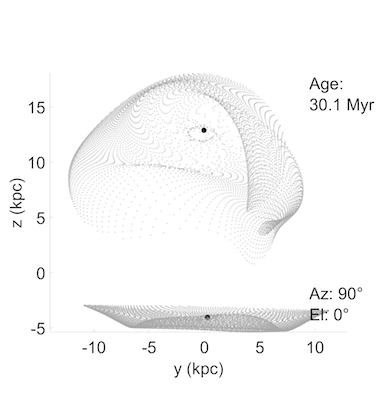}
        }\\ 
    \end{center}
    \caption{Stellar discs for the 500 pc offset collisions, inclinations are a) 20$\degree{}$, b) 45$\degree{}$, c) 65$\degree{}$, d) 90$\degree$.}%
     \label{fig:500pcstellar}
\end{figure*}

\begin{figure*}
     \begin{center}
                \subfigure[]{%
          \includegraphics[width=0.4\textwidth]{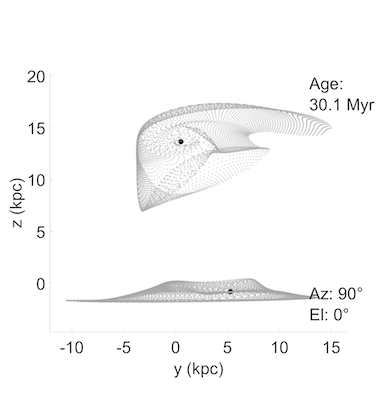}
        }
                \subfigure[]{%
          \includegraphics[width=0.4\textwidth]{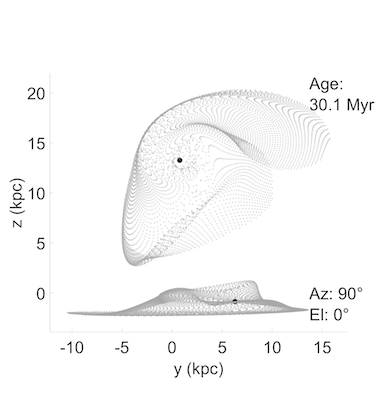}
        }\\ 
                \subfigure[]{%
          \includegraphics[width=0.4\textwidth]{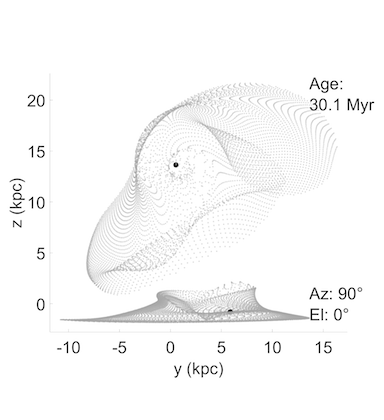}
        }
                \subfigure[]{%
          \includegraphics[width=0.4\textwidth]{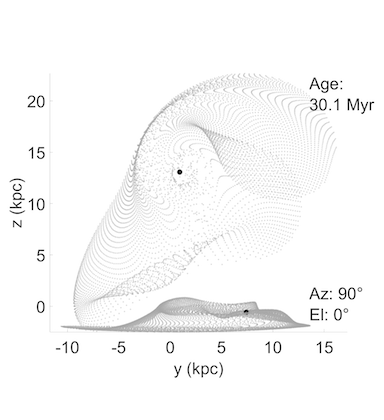}
        }\\ 
    \end{center}
    \caption{Stellar discs for the 10 kpc offset collisions, inclinations are a) 20$\degree{}$, b) 45$\degree{}$, c) 65$\degree{}$, d) 90$\degree$.}%
     \label{fig:10kpcstellar}
\end{figure*}

\indent Stellar discs have been modelled with the same code except with collisions turned off, shown in \cref{fig:500pcstellar,fig:10kpcstellar}.  This is to highlight the gravitational effects on the discs.  Comparing these discs to the splash bridges shows us that in no model does the stellar disc trace out the kinematic evolution of the gas.  In low inclinations we see no stellar component where the splash bridge is located.  For high $90\degree{}$ inclinations G2 does produce large tidal tails, however, they are still displaced by several kpc from where the splash bridge resides.

\section{{Effects of Particle Resolution}}

\begin{figure*}
     \begin{center}
                \subfigure[]{%
          \includegraphics[width=0.4\textwidth]{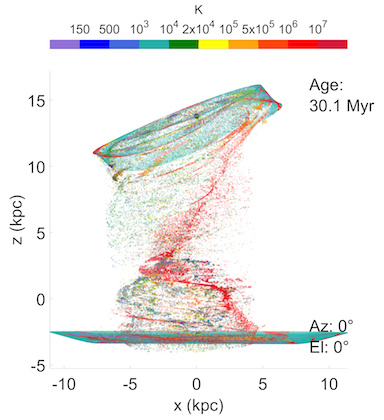}
        }
                \subfigure[]{%
          \includegraphics[width=0.4\textwidth]{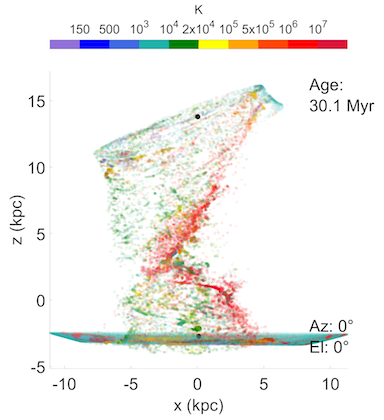}
        }\\ 
                \subfigure[]{%
          \includegraphics[width=0.4\textwidth]{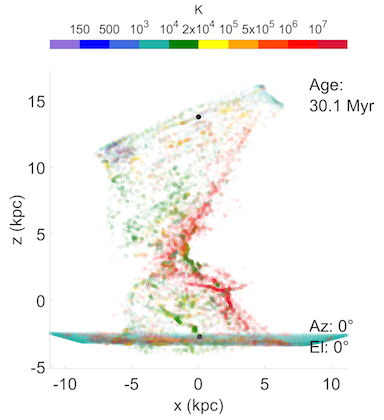}
        }
                \subfigure[]{%
          \includegraphics[width=0.4\textwidth]{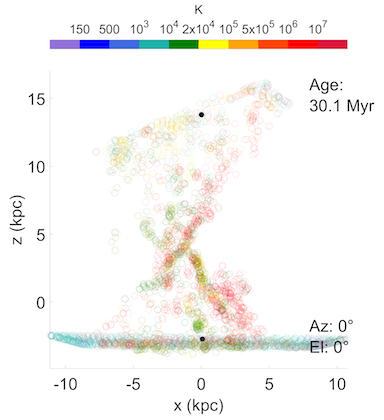}
        }\\ 
    \end{center}
    \caption{Cloud temperature distributions in different resolution (initial cloud size) models all shown at t = 30 Myr.  Resolutions are a) 70 pc, b) 140 pc, c) 200 pc and d) 500 pc.}%
     \label{fig:resolutions1}
\end{figure*}

\begin{figure*}
     \begin{center}
                \subfigure[]{%
          \includegraphics[width=0.4\textwidth]{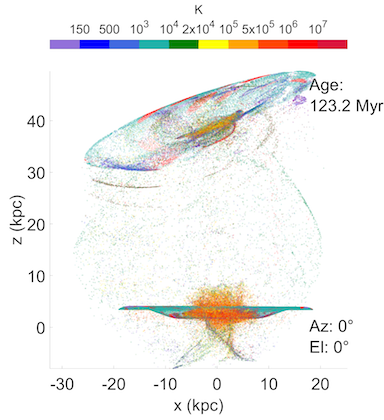}
        }
                \subfigure[]{%
          \includegraphics[width=0.4\textwidth]{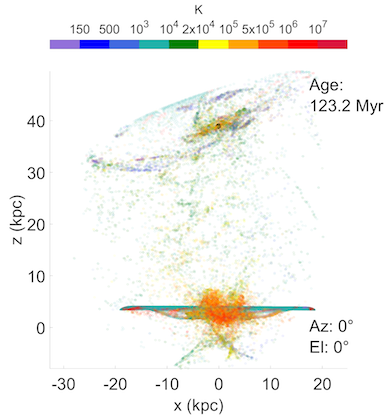}
        }\\ 
                \subfigure[]{%
          \includegraphics[width=0.4\textwidth]{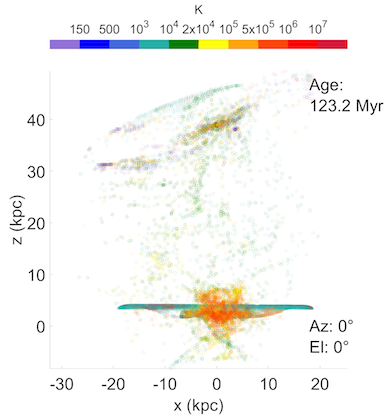}
        }
                \subfigure[]{%
          \includegraphics[width=0.4\textwidth]{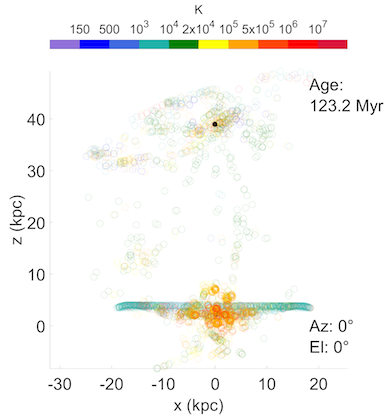}
        }\\ 
    \end{center}
\caption{Cloud temperature distributions in different resolution (initial cloud size) models all shown at t = 123.2 Myr.  Resolutions are a) 70 pc, b) 140 pc, c) 200 pc and d) 500 pc.}%
\label{fig:resolutions2}
\end{figure*}

\indent We have computed four $20\degree{}$ inclination runs with different cloud sizes.  The cloud sizes determine the effective spatial resolution of the runs.  We present snap shots of each resolution at 30 Myr in \cref{fig:resolutions1} and 123.2 Myr in \cref{fig:resolutions2}.  The overall morphology of each bridge is not lost as resolution changes.  The temperature and density histories of the gas clouds in each bridge remain consistent across this range of resolutions as well.  

\bsp    
\label{lastpage}
\end{document}